\documentclass[english,aps,prb,superscriptaddress,reprint,longbibliography]{revtex4-2}
\usepackage[T1]{fontenc}
\usepackage[utf8]{inputenc}
\usepackage{lmodern}
\usepackage{booktabs}
\usepackage{mathrsfs}

\usepackage{amsmath,amssymb,mathtools,bm}
\usepackage{graphicx}
\usepackage{siunitx} 
\usepackage[colorlinks=true,allcolors=blue]{hyperref}

\setcitestyle{numbers,sort&compress} 
\begin{document}

\title{Temperature dependence of spin-model parameters in ferrimagnets}

\author{Marta Yanguas}
  \affiliation{Instituto de Ciencia de Materiales de Madrid, CSIC, Cantoblanco, 28049 Madrid, Spain\looseness=-1}

\author{José M. Lendínez}
  \affiliation{Instituto de Ciencia de Materiales de Madrid, CSIC, Cantoblanco, 28049 Madrid, Spain\looseness=-1}  

\author{Theodor Griepe}
  \affiliation{Instituto de Ciencia de Materiales de Madrid, CSIC, Cantoblanco, 28049 Madrid, Spain\looseness=-1}
  
\author{Rubén M. Otxoa}
  \affiliation{Hitachi Cambridge Laboratory, J. J. Thomson Avenue, Cambridge CB3 0HE, United Kingdom\looseness=-1}
  
\author{Levente Rózsa}
   \affiliation{Department of Theoretical Solid State Physics, Institute of Solid State Physics and Optics,
HUN-REN Wigner Research Centre for Physics, H-1525 Budapest, Hungary}
   \affiliation{Department of Theoretical Physics, Budapest University of Technology and Economics, M\H{u}egyetem rkp. 3, H-1111 Budapest, Hungary}
    
\author{Unai Atxitia}
  \affiliation{Instituto de Ciencia de Materiales de Madrid, CSIC, Cantoblanco, 28049 Madrid, Spain\looseness=-1}

\date{\today}

\begin{abstract}
We compute the temperature dependence of effective spin-model parameters in two-sublattice ferrimagnets with antiparallel and inequivalent sublattice magnetizations using Green's-function theory. The effective intra- and intersublattice exchange interactions, together with the sublattice anisotropies, exhibit distinct thermal renormalizations arising from the different magnetic interactions and correlations of the two sublattices. Although these effective interaction parameters decrease monotonically with increasing temperature, their unequal renormalization produces a strongly nonmonotonic and polarization-dependent long-wavelength spin-wave response. In particular, the spin-wave exchange stiffness exhibits a strongly branch-dependent and nonmonotonic temperature dependence. One polarization develops an enhanced spin-wave stiffness in the vicinity of the angular momentum compensation region despite the monotonic thermal reduction of the underlying exchange interactions. The Green's-function predictions are in quantitative agreement with atomistic spin-dynamics simulations, which provide a numerical benchmark including thermally induced spin correlations within the underlying atomistic spin model. These results establish a direct connection between temperature-dependent atomistic interactions, long-wavelength spin-wave stiffness, and the effective parameters required for finite-temperature micromagnetic and multiscale modeling of ferrimagnetic materials.
\end{abstract}

\maketitle

\section{INTRODUCTION}

Spintronics has emerged as a promising route toward next-generation electronic technologies with improved speed, functionality and energy efficiency~\cite{Zutic2004Spintronics,ManchonRMP2019,Hirohata2020Spintronics}. Over the past decades, significant efforts have been devoted to identifying physical mechanisms and material platforms capable of overcoming the limitations of conventional spintronics based on ferromagnetic materials. As a result, magnetic systems with unconventional ordering and dynamics have attracted growing attention, as they offer new opportunities to tailor spin transport, switching behavior, and information processing~\cite{Gomonay2014,Baltz2018,Barker2021,Kim2022,mejkal2022,Rimmler2025}.

Among these magnetic systems, ferrimagnets—characterized by antiferromagnetically coupled sublattices with unequal magnetic moments—constitute a particularly attractive class for spintronic applications, as they combine advantageous features of both ferromagnets and antiferromagnets. In particular, ferrimagnets can exhibit high-speed domain-wall motion near the compensation temperature \cite{Kim2017, Caretta2020}, as well as ultrafast magnetization dynamics and switching~\cite{Radu2011,Ostler2012,Atxitia2014,Hennecke2019,Jakobs2022,Steinbach_2024,Hennecke2025}. Their spin-wave spectrum, which includes antiferromagnetic-like modes, further makes them promising candidates for terahertz applications \cite{hamara_ultra-high_2024, shen_terahertz_2026, xu_inversion_2026}. Moreover, the finite net magnetization enables conventional electrical control schemes used in ferromagnetic devices~\cite{Zhang2023}, making them promising for device-oriented implementations~\cite{Lalieu2019}.

Crucially, key magnetic properties that enable access to regimes of enhanced dynamical response in ferrimagnets can be tuned via temperature \cite{Schlickeiser2012,barker_two-magnon_2013, donges_unveiling_2020}. Ferrimagnetic materials exhibit a variety of temperature-driven phenomena, including magnetization and angular momentum compensation 
\cite{Kittel1948Ferrimagnetism,Buschow1979Compensation,Kimel2004Ultrafast}. These effects primarily arise from the nonequivalent sublattices, which differ in their exchange interactions, gyromagnetic ratios, and magnetic moments, and therefore respond differently to thermal excitations.

Ferrimagnets are commonly described  using multi-sublattice atomic spin Hamiltonians, typically of Heisenberg type, whose parameters—exchange constants, magnetic anisotropies and sublattice magnetic moments—determine the magnetic ground state and excitation spectrum~\cite{Ostler2011,Garanin2026}.  Nevertheless, realistic modeling of spintronic phenomena in ferrimagnetic devices often requires access to length and time scales that are computationally prohibitive for fully atomistic simulations. In this context, coarse-grained approaches become essential, with thermal effects playing a central role in the dynamical properties. Prominent examples of these models include finite-temperature micromagnetic models, such as those based on the Landau–Lifshitz–Bloch equation, which provides an effective framework for describing both longitudinal and transverse magnetization dynamics at elevated temperatures~\cite{Garanin1997,Chubykalo-Fesenko2006,Atxitia2012,Evans2012, Atxitia2017}. Within coarse-grained models, the effective on-site anisotropy and pairwise interactions parameters, such as exchange and Dzyaloshinskii–Moriya couplings, are intrinsically temperature dependent~\cite{Bastardis2012,Rzsa2017,Evans2020,Rzsa2023}. While the thermal scaling of these effective parameters has been established for ferro- and antiferromagnets, ferrimagnets remain largely unexplored~\cite{Moreno2025}.

The scaling of these effective spin-model parameters can be determined by techniques accounting for thermal fluctuations. Atomistic spin dynamics (ASD) simulations have proven to be a powerful tool to address both equilibrium and non-equilibrium properties of ferrimagnets \cite{Ostler2011, evans_atomistic_2014} \textcolor{black}{and to parameterize multiscale magnetic models. However, their atomistic resolution and the statistical sampling required to obtain accurate finite-temperature properties make them computationally demanding, particularly when a broad temperature range must be explored}. 
Green’s function (GF) methods \textcolor{black}{offer a computationally more efficient, semi-analytical alternative for describing} equilibrium finite-temperature magnetic excitations and spin correlation effects beyond simple mean-field approximations~\cite{Callen,Tyablikov1959,Anderson,FROBRICH2006}. The resulting correlation functions provide direct access to the thermal renormalization and scaling of the effective spin-model parameters.

In this work, we employ GF techniques to systematically investigate the thermal renormalization of effective spin-model parameters in two-sublattice ferrimagnets, including intra- and inter-sublattice exchange interactions and magnetic anisotropies. We validate the GF approach through direct comparison with large-scale atomistic spin-dynamics simulations and analyze how the resulting temperature-dependent parameters determine the long-wavelength spin-wave dynamics, including the zone-center gaps and branch-resolved spin-wave stiffnesses. We further quantify the role of thermal spin correlations through the low-temperature scaling exponents relating the effective exchange interactions to the sublattice magnetizations. 
Altogether, our results establish a consistent framework for connecting finite-temperature atomistic interactions to the effective parameters required in coarse-grained micromagnetic and multiscale descriptions of ferrimagnetic systems.

\section{THEORY}
\label{sec:THEORY}
\begin{figure}[t]
    \centering
    \includegraphics[width=1\linewidth]{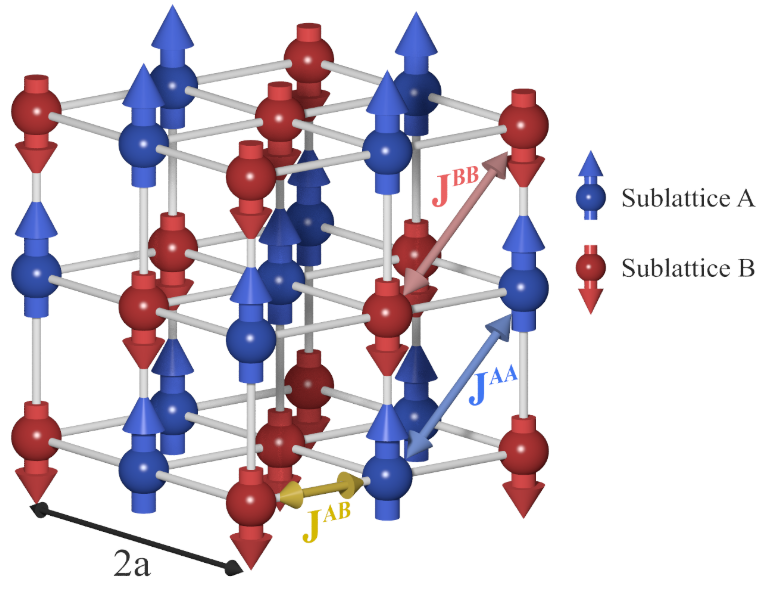}
    \caption{Schematic representation of the ferrimagnetic rock-salt-type lattice considered in this work. The collinear ground state consists of two magnetic sublattices with antiparallel moments aligned along the $z$ axis. Blue and red arrows denote the atomic magnetic moments on sublattices $A$ and $B$, respectively\textcolor{black}{, with their lengths proportional to the corresponding moment magnitudes $\mu_A$ and $\mu_B$}. \textcolor{black}{Exchange interactions are depicted by double-headed arrows: blue for $A$--$A$, red for $B$--$B$, and yellow for $A$--$B$ interactions}}
    \label{fig:rocksalt}
\end{figure}

We consider a two-sublattice ferrimagnet on a rock-salt-type lattice, described by the atomistic spin Hamiltonian
\begin{equation}
    \mathcal{H}= -\frac{1}{2}\sum_{i,j,r,s} J^{rs}_{ij}\mathbf{S}_{ir}\cdot\mathbf{S}_{js} - \sum_{i,r}K_r(S_{ir}^z)^2 
    \label{eq:Hamiltonian}
\end{equation}
where $i,j$ are the lattice site labels and $r,s \in \{A, B\}$ denote the two magnetic sublattices. The variables $\mathbf S_{ir}$ are classical spin vectors of unit length,
$|\mathbf S_{ir}| = 1$, representing the directions of the atomic magnetic moments.
The magnetic moment at site $i$ on sublattice $r$ is given by
$\bm{\mu}_{ir} = \mu_r \mathbf S_{ir}$,
where $\mu_r$ denotes the magnitude of the atomic magnetic moment on sublattice $r$. The first term in Eq. \eqref{eq:Hamiltonian} describes the isotropic Heisenberg exchange interactions, with $J_{ij}^{rs}$ describing the exchange coupling between spins on sites $i$ and $j$ belonging to sublattices $r$ and $s$, respectively. We assume antiferromagnetic nearest-neighbor inter-sublattice exchange interactions \( J^{AB}_{ij} < 0 \), which favor an antiparallel alignment of neighboring spins on sublattices \( A \) and \( B \). We also assume nearest-neighbor intra-sublattice ferromagnetic exchange couplings $J^{AA}_{ij}>0$ and $J^{BB}_{ij}>0$. 
The second term in Eq. \eqref{eq:Hamiltonian} corresponds to the  single-ion uniaxial magnetocrystalline anisotropy, $K_r >0$, on sublattice $r$, which selects the $z$ axis as the easy axis of magnetization. 
As shown schematically in Fig.~\ref{fig:rocksalt}, the classical ground state is collinear: spins on sublattice $A$ point along $+z$, whereas those on sublattice $B$ point along $-z$.

To determine the temperature dependence of the spin-model parameters, we first calculate the spin-wave (SW) dispersion by linearizing the Landau--Lifshitz (LL) equation around the collinear magnetic ground state. The undamped spin dynamics are governed by
\begin{equation}
    \partial_t \mathbf{S}_{ir} = -\gamma_r \mathbf{S}_{ir} \times \mathbf{B}^{\textrm{eff}}_{ir}=
    \frac{\gamma_r}{\mu_r}
    \mathbf{S}_{ir} \times\frac{\partial\mathcal H}{\partial\mathbf S_{ir}},
    \label{equationofmotionLL}
\end{equation}
where $\gamma_r$ represents the sublattice-specific gyromagnetic ratio.

To linearize Eq.~\eqref{equationofmotionLL} around the collinear ground state, we introduce local coordinate frames in which the equilibrium spin direction is $+z$ on both
sublattices. This corresponds to a rotation of the $B$-sublattice spins by $\pi$ about the \(y\) axis, such that
\begin{equation}
\tilde S_{iA}^{\alpha} = S_{iA}^{\alpha},
\qquad
\tilde S_{iB}^{z} = - S_{iB}^{z},
\qquad
\tilde S_{iB}^{\pm} = - S_{iB}^{\mp},
\label{eq:rotation}
\end{equation}
where $\alpha = x,y,z$ and $\tilde S_{ir}^{\pm} = \tilde S_{ir}^x \pm i \tilde S_{ir}^y$ describe the transverse spin components associated with small fluctuations around the equilibrium directions $\tilde S_{ir}^z$. 

By linearizing Eq.~\eqref{equationofmotionLL} and retaining only the first-order terms in the transverse spin components, the different wave vectors $\mathbf{q}$ decouple upon spatial Fourier transformation~\cite{Rzsa2023},
\begin{equation}
    \tilde S_{\mathbf q r}^{\pm}
    =
    \frac{1}{\sqrt{N_r}}
    \sum_{j\in r}
    \tilde S_{jr}^{\pm}
    e^{-i\mathbf q\cdot\mathbf R_j}.
    \label{eq:FT_spin}
\end{equation}
where $N_r$ is the number of spins in sublattice $r$ and $\mathbf R_j$ denotes the lattice position of site $j$.
For the vector
\begin{equation}
    \boldsymbol{\psi}_{\mathbf q}
    =
    \begin{pmatrix}
        \tilde S_{\mathbf q A}^{-}\\
        \tilde S_{\mathbf q B}^{+}
    \end{pmatrix},
\end{equation}
the linearized equation of motion becomes
\begin{equation}
    \omega\,\boldsymbol{\psi}_{\mathbf q}
    =
    \mathcal D_{\mathbf q}\boldsymbol{\psi}_{\mathbf q},
    \label{eq:linearizedequation}
\end{equation}
where $\mathcal{D}_\mathbf{q}$ is the dynamical matrix given by
\begin{equation}
    \mathcal D_{\mathbf q}
    =
    \begin{pmatrix}
        \dfrac{\gamma_A}{\mu_A}\Gamma^{AA}_{\mathbf q}
        &
        \dfrac{\gamma_A}{\mu_A}\Gamma^{AB}_{\mathbf q}
        \\[6pt]
        \dfrac{\gamma_B}{\mu_B}\Gamma^{BA}_{\mathbf q}
        &
        \dfrac{\gamma_B}{\mu_B}\Gamma^{BB}_{\mathbf q}
    \end{pmatrix}.
\end{equation}
One may restrict the analysis to the inter- and intra-sublattice nearest-neighbor exchange interactions in Eq.~\eqref{eq:Hamiltonian}. The corresponding matrix elements are
\begin{align}
\Gamma^{AA}_{\mathbf{q}} &= -2J^{AB}C^{(1)}_{\mathbf{0}} + 2J^{AA} C^{(2)}_{\mathbf{q}} + 2K_A, \label{gammaAA} \\
\Gamma^{AB}_{\mathbf{q}} &=\ \ 2J^{AB} C^{(1)}_{\mathbf{q}}, \label{gammaAB} \\
\Gamma^{BA}_{\mathbf{q}} &= - 2J^{AB} C^{(1)}_{\mathbf{q}}, \label{gammaBA} \\
\Gamma^{BB}_{\mathbf{q}} &= \ \  2J^{AB}C^{(1)}_{\mathbf{0}}   - 2J^{BB} C^{(2)}_{\mathbf{q}} - 2K_B .
\label{gammaBB}
\end{align}
For the rock-salt-type lattice considered here, the structure factors $C^{(n)}_{\mathbf q}$ describe the geometry of the $n$th neighbor shell, with $n=1$ corresponding to nearest-neighbor inter-sublattice couplings and $n=2$ to next-nearest-neighbor intra-sublattice couplings. Explicitly, they are given by
\begin{align}
C^{(1)}_{\mathbf{q}} &= \sum_{\nu} \cos(q_\nu a), \\
C^{(2)}_{\mathbf{q}} &= \sum_{\nu,\kappa \atop \nu \neq \kappa} \Big[ 1 - \cos(q_\nu a) \cos(q_\kappa a) \Big],
\end{align}
where $\nu,\kappa\in\{x,y,z\}$.

The SW spectrum is obtained from the eigenvalues of Eq.~\eqref{eq:linearizedequation}, yielding two SW branches \(\omega^{\pm}_{\mathbf{q}}\):
\begin{align}
  \label{eq:omega_gamma}
     \omega^{\pm}_{\mathbf{q}} =& \frac{1}{2} \left(\frac{\gamma_A}{\mu_A}\Gamma^{AA}_{\mathbf{q}} + \frac{\gamma_B}{\mu_B}\Gamma^{BB}_{\mathbf{q}} \right)\\ \nonumber
      \pm& \frac{1}{2} \sqrt{ \left(\frac{\gamma_A}{\mu_A}\Gamma^{AA}_{\mathbf{q}} - \frac{\gamma_B}{\mu_B}\Gamma^{BB}_{\mathbf{q}}\right)^2 + 4\frac{\gamma_A}{\mu_A}\frac{\gamma_B}{\mu_B} \Gamma^{AB}_{\mathbf{q}} \Gamma^{BA}_{\mathbf{q}} } . 
\end{align}
For the collinear antiparallel state considered here, both SW modes satisfy \(\omega^{+}_{\mathbf{q}} \ge 0\) and \(\omega^{-}_{\mathbf{q}} \le 0\), where the signs indicate the opposite sense of rotation, or chirality, of the two modes. 

\textcolor{black}{At finite temperatures, the algebraic form of Eq.~\eqref{eq:omega_gamma} remains unchanged, while the matrix elements $\Gamma_{\mathbf{q}}^{rs}$ acquire a temperature dependence through the thermally renormalized spin-model parameters. More explicitly, the finite-temperature spectrum is obtained from Eqs.~\eqref{gammaAA}-\eqref{gammaBB} by making the substitutions 
\begin{equation*}
    J^{rs}\rightarrow \mathcal{J}^{rs}(T), \qquad K_r\rightarrow\mathcal{K}_r(T),
\end{equation*}
together with $\mu_r\rightarrow \mu_r m_r(T)$ for the longitudinal sublattice moments, where $m_r(T)$ denotes the reduced sublattice magnetization. Within the GF approach, $\mathcal{J}^{rs}(T)$, $\mathcal{K}_r(T)$, and $m_r(T)$ are determined self-consistently from the temperature-dependent spin correlations and sublattice magnetizations, see Appendix \ref{sec:app_A}. The ASD spin-wave frequencies are obtained as described in Sec.~\ref{sec:simulations}, while the procedure used to extract the corresponding effective parameters is detailed in Appendix~\ref{appC:parameter_extraction}}

\section{SIMULATIONS}
\label{sec:simulations}
 The dynamics of each atomic spin is governed by the stochastic Landau–Lifshitz–Gilbert (sLLG) equation: 
\begin{equation}
        \partial_t \mathbf{S}_{ir} = -\gamma_r \left[\mathbf{S}_{ir} \times \mathbf{B}^{\textrm{eff}}_{ir} -\lambda_r \mathbf{S}_{ir} \times \left(\mathbf{S}_{ir} \times \mathbf{B}^{\textrm{eff}}_{ir}\right)\right], 
    \label{equationofmotionLLG}
\end{equation}
where $i$ labels the spin site and $r$ denotes the magnetic sublattice. The atomic damping parameter used is $\lambda_r=1$, which accelerates the convergence of the statistically averaged quantities.
The effective field is 
\begin{equation}
    \mathbf{B}^{\textrm{eff}}_{ir} = -\frac{1}{\mu_r} \frac{\partial\mathcal{H}}{\partial\mathbf{S}_{ir}}+ \boldsymbol{\zeta}_{ir},
\end{equation}
where $\mathbf{B}^{\textrm{eff}}_{ir}$ includes both the deterministic field derived from the spin Hamiltonian $\mathcal{H}$ and a stochastic field $\boldsymbol{\zeta}_{ir}$ describing thermal fluctuations. \textcolor{black}{The stochastic field is modeled as Gaussian white noise \cite{Atxitia2009} given by}
\begin{equation}
 \langle \zeta_{ir} \rangle = 0 , \hspace{0.5cm}
    \left\langle \zeta_{ir}(0)\,\zeta_{js}(t) \right\rangle
    = \frac{2\,\lambda_r\,k_{\textrm{B}}\,T}{\gamma_r\, \mu_r}\,
      \delta_{ij}\,\delta(t).
      \label{eq:noise}
\end{equation}
The static transverse spin–spin correlation functions are directly related to the SW frequencies $\omega_{\bm{q}}^{\pm}$ of two-sublattice magnets~\cite{Rzsa2023},
\begin{align}
    \left< S^{+}_{-\textbf{q}A} S^{-}_{\textbf{q}A} \right> &
\left< S^{+}_{-\textbf{q}B} S^{-}_{\textbf{q}B} \right>
- \left< S^{+}_{-\textbf{q}B} S^{-}_{\textbf{q}A} \right>
  \left< S^{+}_{-\textbf{q}A} S^{-}_{\textbf{q}B} \right> \nonumber \\
=& 4 \left< S_A^{z} \right>
    \left< S_B^{z} \right>
    \frac{1}{N_c^{2}}
    \frac{\gamma_{A}}{\mu_A}
    \frac{\gamma_{B}}{\mu_B}
    \frac{(k_{\textrm{B}} T)^2}{\omega_\textbf{q}^{+}\omega_\textbf{q}^{-}} ,
    \label{omegaprod}
\end{align}
\begin{equation}
\frac{\left\langle S^{+}_{-\mathbf{q}A} S^{-}_{\mathbf{q}A} \right\rangle}
     {2\gamma_{A} \mu_A^{-1} \left\langle S_A^{z} \right\rangle}
+ \frac{\left\langle S^{+}_{-\mathbf{q}B} S^{-}_{\mathbf{q}B} \right\rangle}
       {2\gamma_{B} \mu_B^{-1} \left\langle S_B^{z} \right\rangle}
= \frac{1}{N_c}\,
  \frac{k_{\textrm{B}} T \left(\omega_\mathbf{q}^{+} + \omega_{\mathbf{q}}^{-}\right)}
       {\omega_\mathbf{q}^{+}\omega_{\mathbf{q}}^{-}},
       \label{omegasum}
\end{equation}
where $N_c$ denotes the number of spins per sublattice, $k_{\textrm{B}}$ is Boltzmann's constant, and $T$ is the temperature of the system. 

The product of the frequencies $\omega_{\textrm{prod}}= \omega_{\mathbf{q}}^+\omega_{\mathbf{q}}^-$ is given by Eq. \eqref{omegaprod}, which also yields the sum $\omega_{\textrm{sum}}= \omega_{\mathbf{q}}^+ + \omega_{\mathbf{q}}^-$ from Eq. \eqref{omegasum}. The individual frequencies can be calculated as 
\begin{equation}
    \omega_{\mathbf{q}}^{\pm} = \frac{1}{2}\left[\omega_{\textrm{sum}} \pm \sqrt{\omega_{\textrm{sum}}^2 -4 \omega_{\textrm{prod}}}\right].
\end{equation}
The spin-spin correlation functions in Eqs.~\eqref{omegaprod} and \eqref{omegasum} can be calculated from ASD simulations by performing the spatial Fourier transform of the real-space spin configurations. Long simulation times of 5~ns, with averaging every 2~fs, are required in the simulations performed here to ensure accurate evaluation of the expectation values.  Three-dimensional rock-salt-type structures consisting of a $64\times64\times64$ spin grid with periodic boundary conditions in all directions were used during the simulations. 

\section{RESULTS}
\begin{figure}
    \centering
    \includegraphics[width=\linewidth]{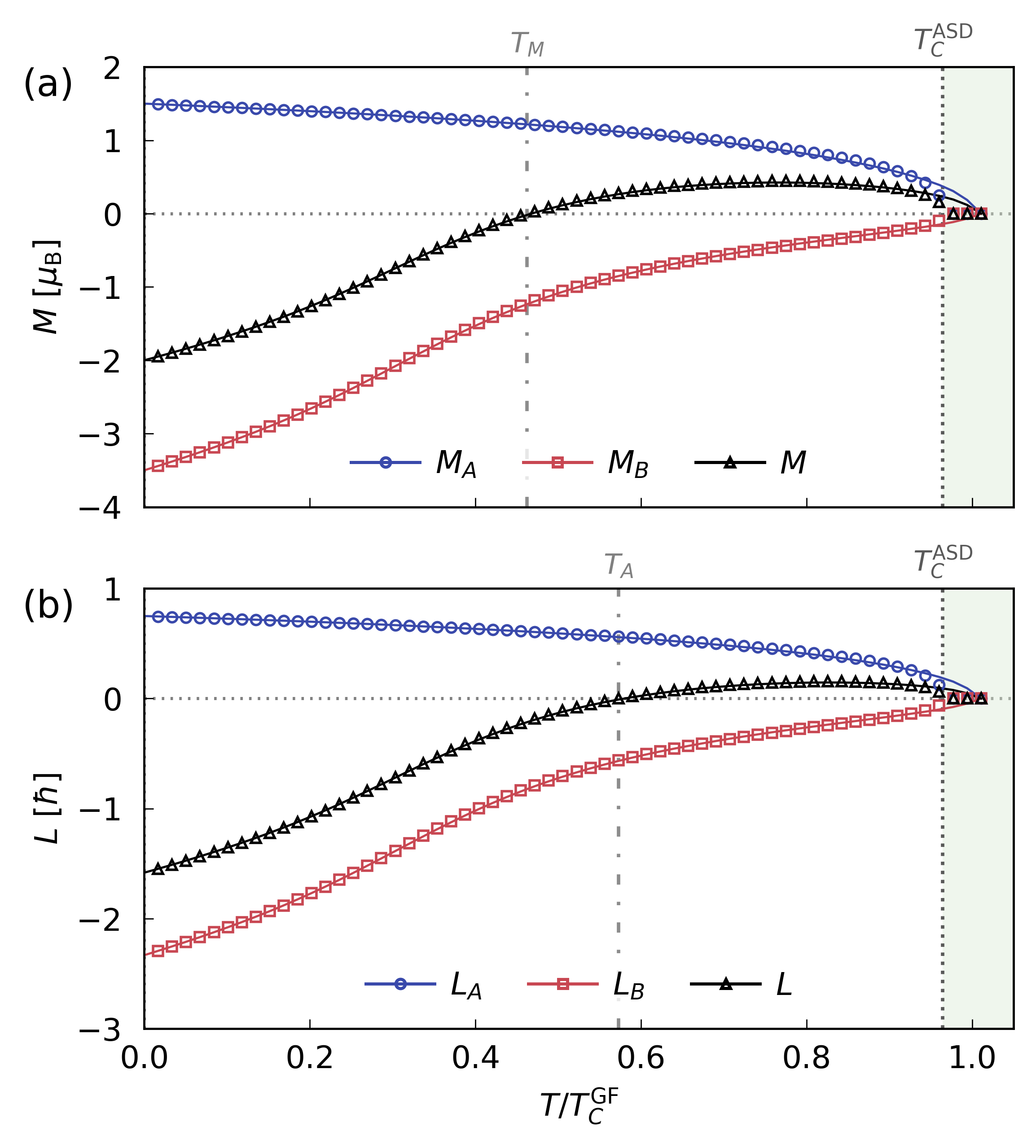}
    \caption{
    Temperature dependence of (a) the signed sublattice magnetization $M_A=\mu_A m_A$ and $M_B=-\mu_B m_B$, together with the net magnetization $M=M_A+M_B$, and (b) the corresponding sublattice angular momenta $L_r=M_r/\gamma_r$ and their sum \(L=L_A+L_B\). Symbols denote ASD results and solid lines show GF calculations. Temperatures are normalized by the GF critical temperature $T_C^{\mathrm{GF}}$. Vertical dashed lines indicate the magnetization compensation temperature $T_M$, the angular momentum compensation temperature $T_A$, and the ASD critical temperature $T_C^{\mathrm{ASD}}$. The shaded region corresponds to $T_C^{\mathrm{ASD}}<T$. The atomistic spin-model parameters are listed in Table~\ref{tab:model_parameters}.
    }
    \label{fig:mag}
\end{figure}

\subsection{Magnetizations and angular momenta}

We consider a representative two-sublattice ferrimagnet with atomistic spin-model parameters characteristic of rare-earth transition-metal ferrimagnets, such as GdFe alloys \cite{Ostler2011,Radu2011,Ostler2012,Jakobs2022}. The parameters, summarized in Table~\ref{tab:model_parameters}, are used throughout both the GF calculations and ASD simulations.

As shown in Fig.~\ref{fig:mag}(a), a characteristic thermal feature of these ferrimagnets is the distinct temperature dependence of the sublattice magnetizations, $M_A=\mu_A m_A$ and $M_B=-\mu_B m_B$, where $m_r$ denotes the magnitude of the reduced sublattice magnetization. This inequivalent thermal behavior arises from the different balance between intra- and inter-sublattice exchange interactions on the two sublattices. In particular, the substantially stronger intra-sublattice exchange within sublattice $A$ stabilizes its magnetic order more effectively against thermal fluctuations, whereas the weaker intra-sublattice coupling on sublattice $B$ leads to a stronger influence of the inter-sublattice exchange and a more rapid reduction of $M_B$ with increasing temperature. Consequently, the two sublattice magnetizations exhibit markedly different thermal evolutions.

\begin{table}[t!]
    \centering
    \setlength{\belowcaptionskip}{6pt}
    \renewcommand{\arraystretch}{1.15}
    \caption{
    Atomistic spin-model parameters of the two-sublattice ferrimagnet considered in the calculations. Here, $   \gamma_e=1.75\times10^{11}\,\mathrm{T^{-1}s^{-1}}$ is the electron's gyromagnetic ratio, and $\mu_{\mathrm B}$ denotes the Bohr magneton.
    }
    \label{tab:model_parameters}
    \begin{tabular*}{\columnwidth}{@{\extracolsep{\fill}}lcc@{}}
        \toprule
        & Sublattice \(A\) & Sublattice \(B\) \\
        \midrule
        \(J^{rr}\,(\mathrm{meV})\)     & 15 & 3 \\
        \(K_r\,(\mathrm{meV})\)        & 1  & 1 \\
        \(\mu_r/\mu_{\mathrm B}\)      & 1.5  & 3.5 \\
        \(\gamma_r/\gamma_e\)          & 1    & 0.75 \\
        \midrule
        \(J^{AB}\,(\mathrm{meV})\) & \multicolumn{2}{c}{-3} \\
        \bottomrule
    \end{tabular*}
\end{table}

For the present parameter set, the unequal atomic moments and distinct thermal evolution of the two sublattices give rise to a magnetization compensation temperature $T_M$, where the total magnetization $M=M_A+M_B$ vanishes, and an angular momentum compensation temperature $T_A$, where the total angular momentum vanishes (see Fig.~\ref{fig:mag}(b)). These two temperatures are different due to the different gyromagnetic ratios on the two sublattices. In the interval $T_M<T<T_A$, the net magnetic moment and net angular momentum have opposite signs. Consequently, the coupling to a uniform magnetic field is governed by $M$, whereas the sign of the gyrotropic response is determined by $L$. The separation between $T_M$ and $T_A$ is a characteristic ferrimagnetic effect and underlies the enhanced dynamical response near angular momentum compensation \cite{Kim2017,Schlickeiser2012}.

The ASD and GF results are in quantitative agreement over a broad temperature range. The independently estimated critical temperatures satisfy $T_C^{\mathrm{ASD}} \simeq 0.97\,T_C^{\mathrm{GF}}$, corresponding to a relative difference of approximately $3\%$. This small deviation may arise from the decoupling approximation inherent to the GF treatment, finite-size and sampling effects in the atomistic simulations. Similar method-dependent deviations and finite-size effects near the ordering temperature have been reported in previous comparisons of GF approaches with numerical spin simulations \cite{Bastardis2012,Rzsa2023}. Moreover, close to $T_C^{\mathrm{ASD}}$, the magnetization cannot be determined accurately due to finite-size effects and limited simulation times, as the system begins to switch between the $+z$ and $-z$ orientations. Therefore, small deviations between the ASD and GF results in this narrow temperature range should not be attributed solely to a breakdown of the GF description.

\subsection{Spin-wave spectrum and effective parameters}

In addition to the temperature dependence of the sublattice magnetizations and angular momenta, we characterize the thermal renormalization of the magnetic interactions through the effective exchange parameters, $\mathcal{J}^{rs}(T)$, and on-site anisotropies, $\mathcal{K}_{r}(T)$. These quantities govern the low-energy SW dynamics and provide the natural link between the atomistic model and a finite-temperature micromagnetic description~\cite{Atxitia2017}. 
Their renormalization is directly reflected in the SW spectrum, particularly in the zone-center gaps and long-wavelength curvature. We therefore first examine the temperature evolution of the SW spectrum and subsequently analyze the effective interaction parameters responsible for its renormalization.

\begin{figure}
    \centering
\includegraphics[width=1\linewidth]{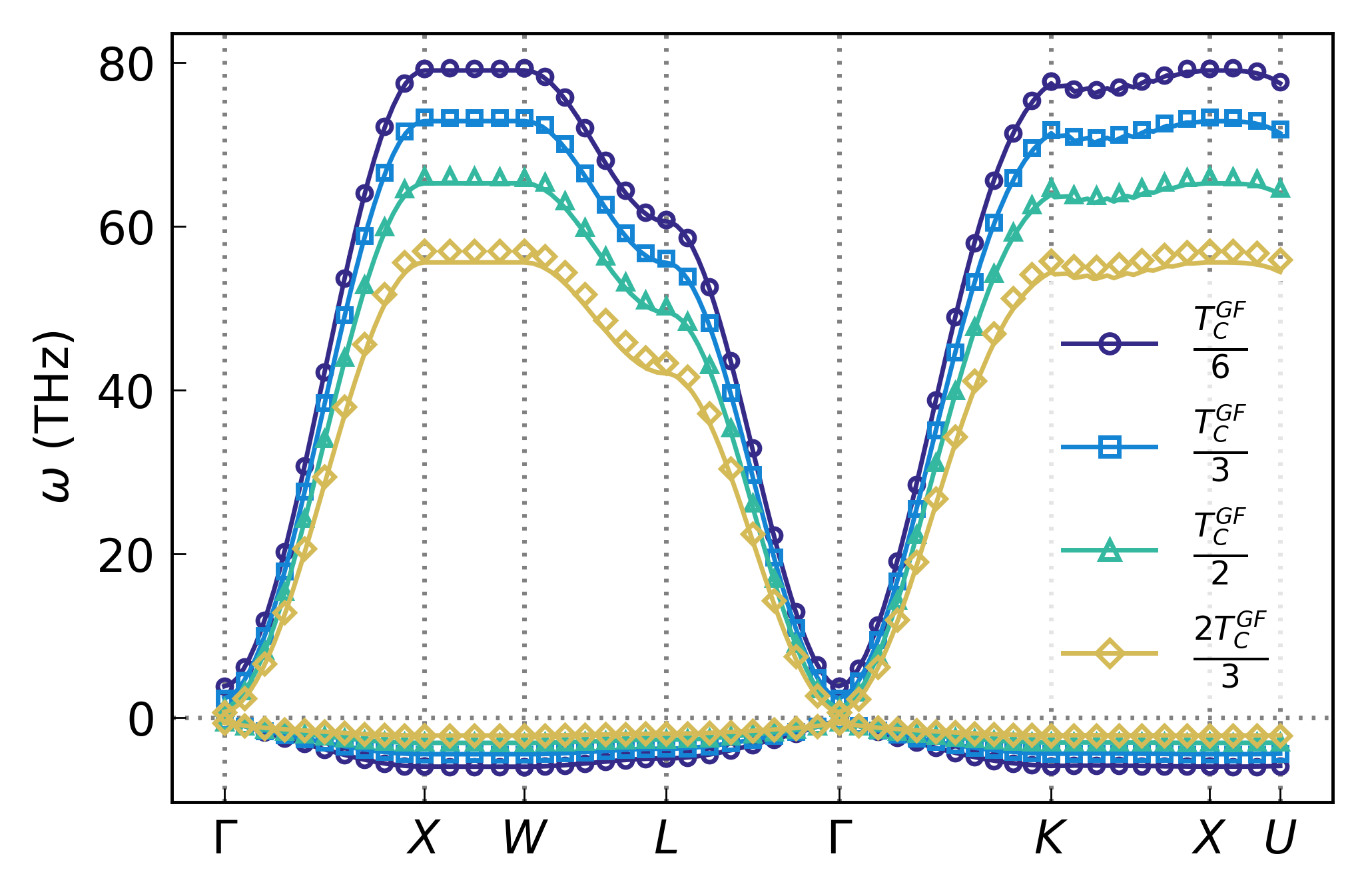}
\caption{
    Temperature evolution of the spin-wave spectrum of the two-sublattice ferrimagnet along the high-symmetry path $\Gamma\!-\!X\!-\!W\!-\!L\!-\!\Gamma\!-\!K\!-\!X\!-\!U$ of the first Brillouin zone of the rock-salt-type structure. Results are shown at $T/T_C^{\mathrm{GF}}=1/6,\,1/3,\,1/2,$ and $2/3$. Symbols denote frequencies extracted from atomistic spin-dynamics simulations, while solid lines show 
    GF calculations. The atomistic spin-model parameters are listed in Table~\ref{tab:model_parameters}.
    }
    \label{fig:sw}
\end{figure}

Figure~\ref{fig:sw} shows the SW spectrum along the high-symmetry direction $\Gamma\!-\!X\!-\!W\!-\!L\!-\!\Gamma\!-\!K\!-\!X\!-\!U$ of the face-centered-cubic Brillouin zone for $T/T_C^{\rm GF}=1/6$, $1/3$, $1/2$, and $2/3$, covering a broad range of the magnetically ordered phase. The frequencies obtained from ASD simulations are represented by symbols, while the solid lines correspond to the GF calculations.

The spectrum consists of two branches with opposite frequency signs, $\omega_{\mathbf q}^{+}>0$ and $\omega_{\mathbf q}^{-}<0$, corresponding to SW modes with opposite circular polarizations. Over most of the Brillouin zone, increasing temperature leads to a progressive reduction in the magnitude of the SW frequencies and a concomitant narrowing of the spectral bandwidth. This thermal softening is particularly pronounced for the positive-frequency branch, whose maxima in the $X\!-\!W$ and $K\!-\!X\!-\!U$ regions decrease substantially with increasing temperature. The negative-frequency branch also shifts toward zero frequency, although its characteristic energy scale remains considerably smaller. Thus, away from the zone center, the dominant finite-temperature effect is a systematic softening of both SW branches.

The behavior near $\Gamma$ is more subtle. In the long-wavelength limit, the zone-center gaps and dispersion curvatures exhibit a temperature dependence that differs from the monotonic softening observed at larger wave vectors. \textcolor{black}{The expansion $\omega_{\mathbf q}^{\pm}=\omega_0^\pm+\mathcal A^\pm(qa_0)^2+\mathcal O[(qa_0)^4]$, derived in Appendix~\ref{sec:appB}, shows that $\omega_0^\pm$ and $\mathcal A^\pm$ depend on different combinations of the effective exchange and anisotropy parameters.} Their evolution therefore provides a sensitive probe of the unequal thermal renormalization of the different magnetic interactions. In particular, one circularly polarized branch develops a pronounced nonmonotonic temperature dependence, which is analyzed in detail below.

The GF theory reproduces both the overall shape of the SW spectrum and its temperature evolution over the entire Brillouin-zone path. \textcolor{black}{Small systematic deviations appear at the highest temperature considered, where the GF frequencies slightly underestimate the magnitude of the ASD results, particularly in the relatively flat regions of the positive-frequency branch. The tendency for the discrepancy to increase with temperature is consistent with the reduced accuracy of the correlation-decoupling approximation in the critical regime, where fluctuations become increasingly important. Nevertheless, the agreement remains very good over most of the ordered phase, demonstrating} that the GF formalism captures the dominant finite-temperature renormalization of both circularly polarized modes, including the long-wavelength features relevant to the effective gap and exchange stiffness.

To identify the microscopic origin of these spectral changes, we next determine the temperature-dependent exchange interactions and on-site anisotropies. Within the ASD approach, \textcolor{black}{the SW spectrum at each temperature is fitted to the finite-temperature form of the atomistic dispersion in Eq.~\eqref{eq:omega_gamma}. The longitudinal sublattice moments entering the dispersion are taken as $\mu_r m_r(T)$, with $m_r(T)$ obtained directly from the ASD simulations, while the exchange and anisotropy parameters are treated as effective temperature-dependent quantities, $\mathcal{J}^{rs}(T)$ and $\mathcal{K}_{r}(T)$}. The exchange parameters are fitted primarily to the finite-wave-vector part of the spectrum, which determines the overall dispersion and bandwidth, whereas the anisotropies are extracted from the small wave-vector region with the exchange parameters held fixed. \textcolor{black}{Further details of the fitting procedure are provided in Appendix~\ref{appC:parameter_extraction}.} By contrast, \textcolor{black}{within the GF approach, both the finite-temperature SW spectrum and the effective interaction parameters are obtained self-consistently from the spin correlation functions, following the formalism of Ref.~\cite{Rzsa2023}}.

\begin{figure}
    \centering
    \includegraphics[width=1\linewidth]{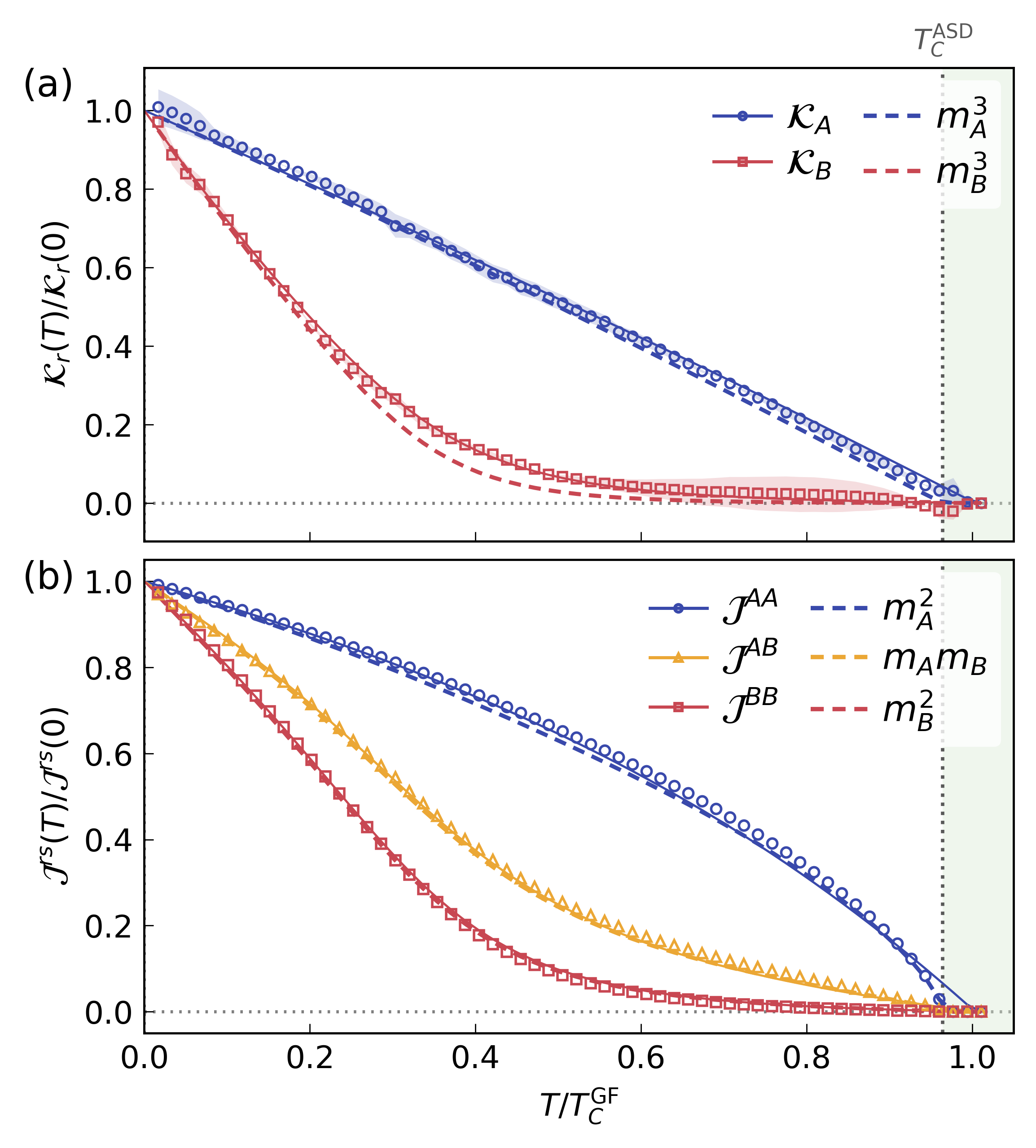}
    \caption{Temperature dependence of (a) the effective single-ion anisotropy parameters $\mathcal{K}_A(T)$ and $\mathcal{K}_B(T)$ and (b) the effective intra- and inter-sublattice exchange parameters $\mathcal{J}^{AA}(T)$, $\mathcal{J}^{AB}(T)$, and $\mathcal{J}^{BB}(T)$, all normalized by their zero-temperature values. Symbols denote parameters extracted from the ASD spin-wave spectra, while solid lines show the 
    GF predictions. In panel (a), the shaded bands indicate the sensitivity of the ASD-extracted anisotropies to the exchange parameters held fixed during the small wave-vector fits. The fitting uncertainties of the exchange parameters in panel (b) are smaller than the symbol size. Dashed curves show the uncorrelated reference scalings $\mathcal{J}^{AA}/\mathcal{J}^{AA}(0)=m_A^2$, $\mathcal{J}^{AB}/\mathcal{J}^{AB}(0)=m_A m_B$, and $\mathcal{J}^{BB}/\mathcal{J}^{BB}(0)=m_B^2$, together with the single-ion Callen-Callen reference scaling $\mathcal{K}_r/\mathcal{K}_r(0)=m_r^3$. Temperatures are normalized by $T_C^{\mathrm{GF}}$, and the shaded temperature region corresponds to $T_C^{\mathrm{ASD}}<T$. The atomistic spin-model parameters are listed in Table~\ref{tab:model_parameters}.}
    \label{fig:K_J}
\end{figure}

The resulting temperature dependences are shown in Fig.~\ref{fig:K_J}. Figure~\ref{fig:K_J}(a) shows the effective on-site anisotropies, while Fig.~\ref{fig:K_J}(b) presents the effective intra- and intersublattice exchange interactions. All quantities are normalized by their zero-temperature values. The exchange parameters extracted from the ASD spectra are robust, with fitting uncertainties smaller than the symbol size. By contrast, the anisotropy parameters are more sensitive to the exchange values held fixed during the low-wave-vector fits. The shaded bands in Fig.~\ref{fig:K_J}(a) quantify this sensitivity and span the range obtained when the exchange parameters are fixed either to the ASD-extracted values or to the GF predictions. Further details of the fitting procedure and uncertainty estimation are provided in Appendix~\ref{appC:parameter_extraction}.

For comparison, the dashed curves in Fig.~\ref{fig:K_J} show the uncorrelated reference scalings expected from the reduced sublattice magnetizations: $\mathcal{J}^{AA}(T)/\mathcal{J}^{AA}(0)=m_A^2$, $\mathcal{J}^{AB}(T)/\mathcal{J}^{AB}(0)=m_A m_B$, $\mathcal{J}^{BB}(T)/\mathcal{J}^{BB}(0)=m_B^2$, and $\mathcal{K}_r(T)/\mathcal{K}_r(0)=m_r^3$. These relations provide a simple baseline against which the effect of thermal spin correlations can be assessed. The effective exchange interactions obtained from both ASD and GF calculations generally decrease more slowly than the corresponding uncorrelated reference curves, indicating that transverse spin correlations reduce the rate of thermal suppression. This correlation-induced renormalization also influences the long-wavelength curvature of the SW spectrum and hence the temperature dependence of the spin-wave exchange stiffness.

The effective anisotropies shown in Fig.~\ref{fig:K_J}(a) exhibit a substantially stronger relative thermal reduction than the exchange interactions in Fig.~\ref{fig:K_J}(b). In particular, $\mathcal{K}_B(T)$ decreases more rapidly than $\mathcal{K}_A(T)$, consistent with the faster loss of magnetic order on sublattice $B$. Both anisotropies approximately follow the single-ion Callen--Callen scaling, $\mathcal{K}_r(T)\propto m_r^3(T)$, reasonably well at low temperatures ~\cite{CallenCallen1966}. The different thermal evolutions of $\mathcal{K}_A(T)$ and $\mathcal{K}_B(T)$ therefore arise primarily from the inequivalent temperature dependences of the two sublattice magnetizations. As shown by the long-wavelength expansion of the SW spectrum presented in Appendix~\ref{sec:appB}, the zone-center SW energies depend on both exchange and anisotropy contributions, the distinct renormalization of these quantities provides a natural explanation for the nontrivial temperature dependence of the SW gap near $\Gamma$ discussed in Sec.~\ref{sec:longwave} below. Overall, the ASD-extracted interaction parameters and the GF predictions are in quantitative agreement throughout most of the ordered phase, demonstrating that the GF theory captures the distinct thermal renormalization of the intra- and inter-sublattice exchange interactions and anisotropies.

\begin{figure}
    \centering
    \includegraphics[width=\linewidth]{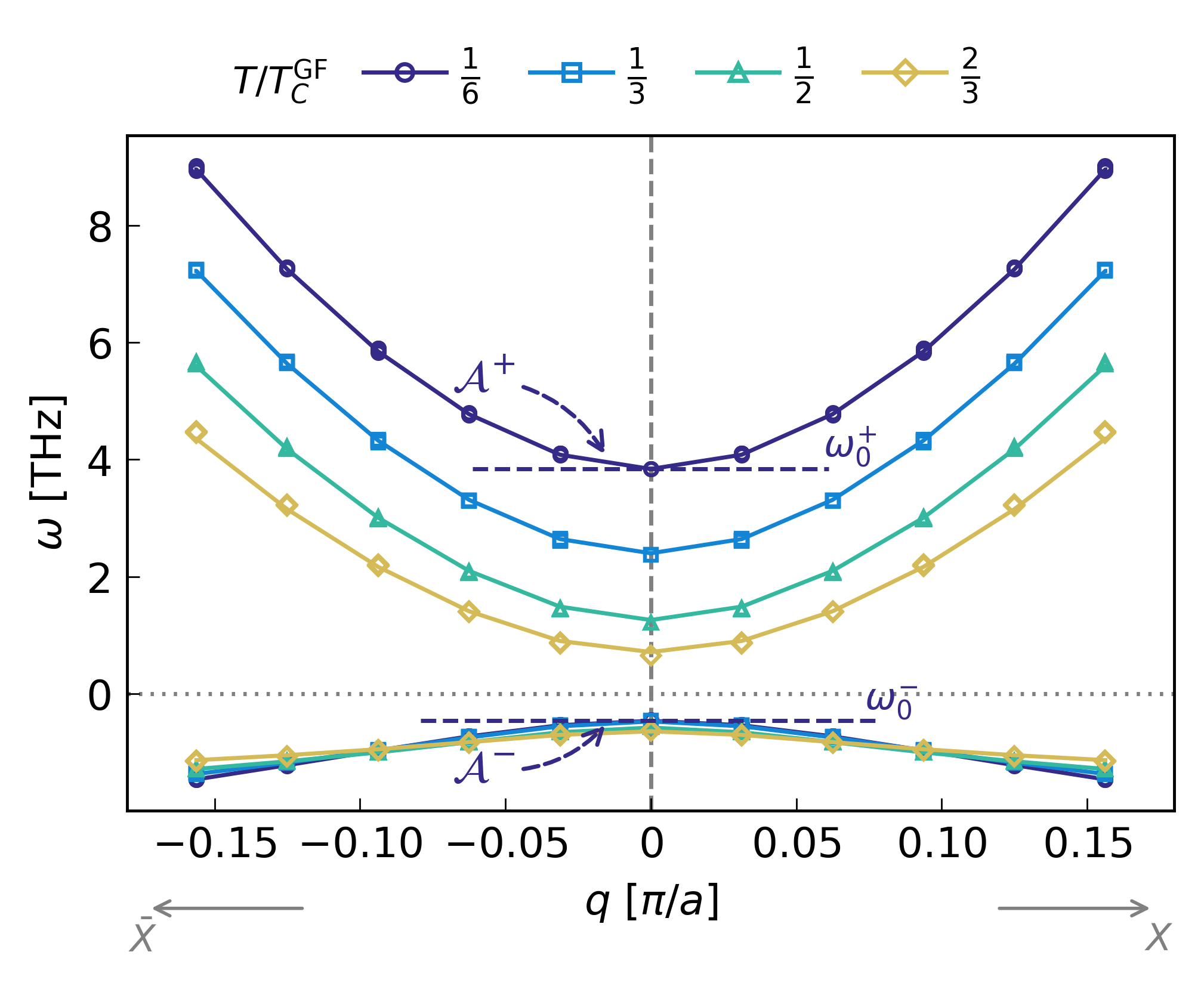}
    \caption{Spin-wave dispersion near the Brillouin-zone center along the $\Gamma X$ direction for $T/T_C^{\mathrm{GF}}=1/6$, $1/3$, $1/2$, and $2/3$. Symbols denote frequencies extracted from ASD simulations, while solid lines show GF calculations. The zone-center gaps $|\omega_0^\pm(T)|$  and branch-resolved curvature coefficients $|\mathcal A^\pm(T)|$, defined through the long-wavelength expansion of the spin-wave spectrum, are indicated schematically. The atomistic spin-model parameters are listed in Table~\ref{tab:model_parameters}.
    }
    \label{fig:sw-zoom}
\end{figure}

\begin{figure}
    \centering
    \includegraphics[width=\linewidth]{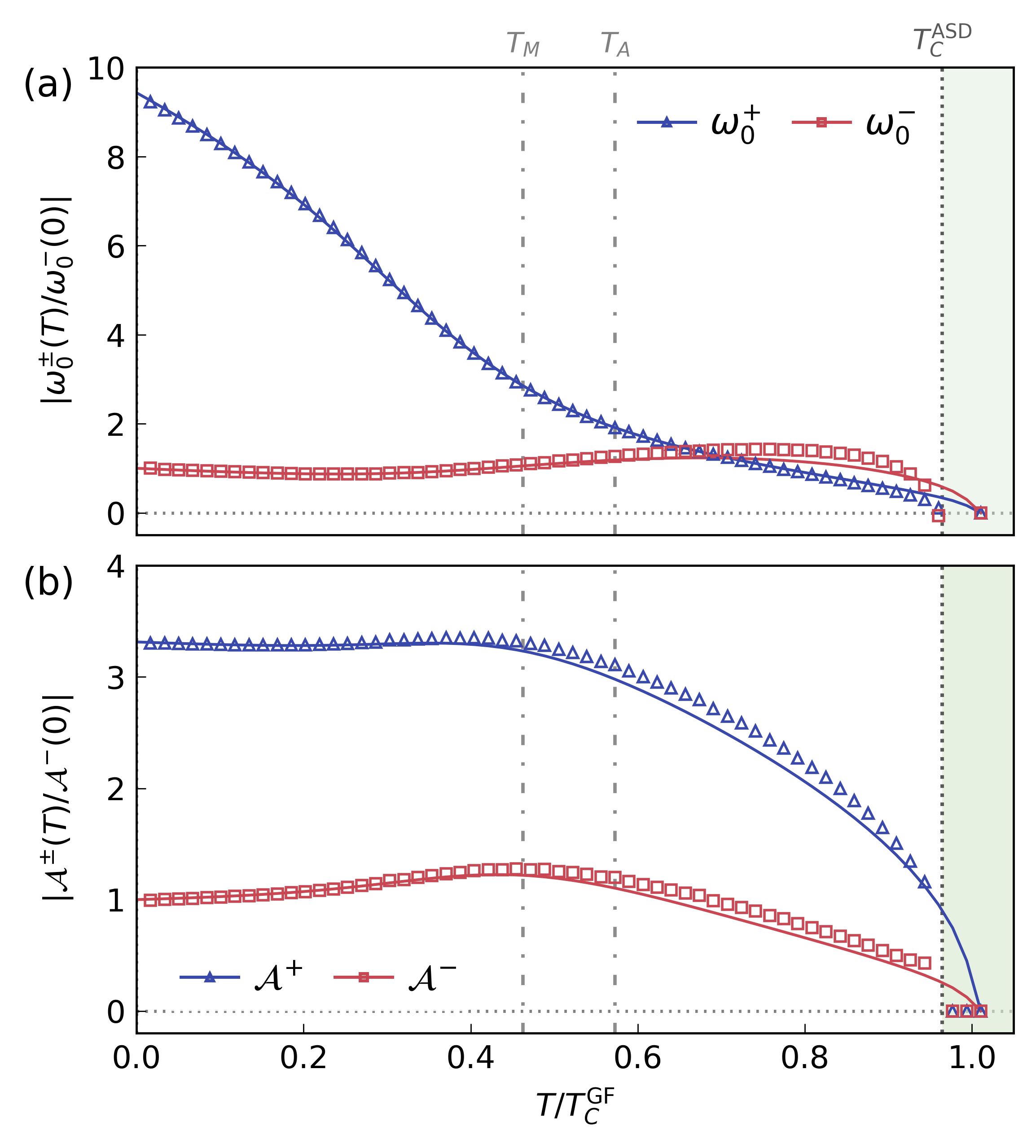}
    \caption{Long-wavelength spin-wave dynamics of the two-sublattice ferrimagnet. (a) Magnitudes of the zone-center gaps, $|\omega_0^\pm(T)|$, normalized by $|\omega_0^-(0)|$. (b) Magnitudes of the branch-resolved spin-wave curvature coefficients, $|\mathcal A^\pm(T)|$, normalized by $|\mathcal A^-(0)|$. Symbols denote quantities evaluated using the parameters extracted from the ASD spectra, while solid lines show the 
    GF predictions. Vertical dash-dotted lines indicate the magnetization-compensation temperature $T_M$, the angular-momentum-compensation temperature $T_A$, while the vertical dotted line indicates the ASD critical temperature $T_C^{\mathrm{ASD}}$. The shaded region corresponds to $T_C^{\mathrm{ASD}}<T$. The atomistic spin-model parameters are listed in Table~\ref{tab:model_parameters}.
    }
    \label{fig:gap_stiffness}
\end{figure}

\subsection{Long-wavelength spin-wave dynamics\label{sec:longwave}}

The unequal thermal renormalization of the effective exchange interactions and uniaxial anisotropies has particularly pronounced consequences in the long-wavelength limit. Figure~\ref{fig:sw-zoom} shows the SW dispersion around the Brillouin-zone center at different temperatures.
For \(qa\ll1\), both branches are accurately described by the quadratic expansion
\begin{equation}
\omega_{\mathbf q}^{\pm}(T)
=\omega_0^\pm(T)+\mathcal A^\pm(T)(qa)^2+\mathcal O\!\left[(qa)^4\right], 
\end{equation}
where $a$ is the lattice constant and $q=|\mathbf{q}|$. \textcolor{black}{Here, \(\omega_0^\pm(T)\) are the signed zone-center frequencies, whose magnitudes \(|\omega_0^\pm(T)|\) define the corresponding SW gaps, while \(\mathcal A^\pm(T)\) are the branch-resolved SW curvature coefficients, directly related to the long-wavelength spin-wave stiffness.} Analytical expressions for these quantities are derived in Appendix~B, following the general long-wavelength framework of Ref.~\cite{Lendinez2026}. At each temperature, these expressions are evaluated independently using either the parameters extracted from the ASD spectra or the GF parameters shown in Fig.~\ref{fig:K_J}.

Figure~\ref{fig:gap_stiffness}(a) reveals markedly different temperature dependences of the two zone-center gaps. The magnitude of the positive-frequency gap, $|\omega_0^+|$, decreases monotonically with increasing temperature, broadly following the overall thermal softening of the SW spectrum. In contrast, $|\omega_0^-|$ exhibits a pronounced nonmonotonic behavior: it increases over an intermediate temperature range before eventually decreasing toward zero as the magnetic ordering transition is approached. This behavior shows that the long-wavelength dynamics cannot be inferred simply from the monotonic decrease of the individual exchange and anisotropy parameters. Instead, the gap is determined by their combined contributions together with the temperature-dependent magnetic state of the two sublattices, whose unequal thermal evolution modifies the balance between the two ferrimagnetic eigenmodes.

For the present parameter set, the two gap magnitudes become nearly equal in the vicinity of the angular-momentum-compensation temperature $T_A$. This near-equality is consistent with the reduced dynamical asymmetry between the sublattices near angular-momentum-compensation, where the oppositely polarized magnon branches approach each other. A related reduction of the magnon-band separation near compensation has been discussed previously in the context of two-magnon-mediated ferrimagnetic dynamics~\cite{barker_two-magnon_2013}. Below this temperature region, the positive-frequency branch has the larger gap magnitude, whereas the negative-frequency branch corresponds to the lower-frequency excitation. Across the compensation region, this ordering is reversed, reflecting the change in which sublattice dominates the net angular momentum while both sublattices remain magnetically ordered.

An even more pronounced nonmonotonic behavior is observed in the SW curvature coefficients shown in Fig.~\ref{fig:gap_stiffness}(b). The magnitude $|\mathcal A^+|$ remains approximately constant at low temperatures and decreases upon further heating. By contrast, $|\mathcal A^-|$ increases substantially with temperature, reaches a pronounced maximum in the vicinity of the compensation region, and decreases only at higher temperatures as the ordering transition is approached. Hence, despite the monotonic thermal reduction of all underlying exchange interactions, the long-wavelength dispersion of one circularly polarized mode becomes progressively steeper over a finite temperature interval. This behavior emphasizes that the branch-resolved curvature is governed not simply by the magnitude of the effective exchange parameters, but by their interplay with the temperature-dependent sublattice magnetizations entering the collective ferrimagnetic dynamics.

\textcolor{black}{The corresponding group velocity in the long-wavelength limit is obtained from the slope of the SW dispersion,}
\begin{equation}
   \frac{\partial \omega_{\mathbf q}^\pm}{\partial q}
\simeq
2\mathcal A^\pm(T)a^2q, 
\end{equation}
Thus, at a fixed small wave vector, an increase in $|\mathcal A^\pm|$ corresponds to an increase in the magnitude of the group velocity. The pronounced growth of $|\mathcal A^-|$ therefore implies faster propagation of this circularly polarized mode within the long-wavelength regime around the compensation temperature, while the curvature of the opposite branch is progressively reduced. The resulting dynamics are thus strongly temperature and polarization dependent. We note, however, that the SW propagation length is determined not only by the group velocity but also by the mode lifetime and damping, and therefore cannot be inferred from the dispersion alone.

These results highlight an important distinction between the thermal evolution of the effective interaction parameters and that of the collective long-wavelength dynamics. Although 
the effective spin-model parameters all decrease monotonically with increasing temperature, the SW gaps and curvatures depend on specific combinations of these quantities together with the temperature-dependent sublattice magnetizations. Their competing thermal dependences can therefore generate strongly nonmonotonic collective properties, particularly in the vicinity of angular momentum compensation. This demonstrates why the full temperature dependence of the effective spin-model parameters must be retained when constructing a quantitative finite-temperature description of ferrimagnetic spin dynamics.

\subsection{Low-temperature scaling}
\label{Low-temperature scaling}
The temperature dependence of the effective spin-model parameters can be conveniently expressed in terms of the reduced sublattice magnetizations rather than the bath temperature. Such a representation is particularly useful for finite-temperature micromagnetic and nonequilibrium models, where the instantaneous magnetic order provides a more direct measure of the state of the spin system. It also allows the thermal renormalization of the effective interactions to be characterized independently of the specific temperature dependence of the sublattice magnetizations.

\textcolor{black}{Within the mean-field approximation, where spin correlations are neglected,} the effective exchange parameters follow the reference scalings $\mathcal J^{AA}\propto m_A^2$, $\mathcal J^{BB}\propto m_B^2$, and $\mathcal J^{AB}\propto m_A m_B$, corresponding to the dashed curves in Fig.~\ref{fig:K_J}(b). As discussed in the previous section, the effective exchange interactions obtained from both ASD and GF calculations decrease more slowly than these uncorrelated parameters, reflecting the influence of transverse spin correlations. At low temperatures, these correlation effects can be incorporated through the power-law relations
\begin{align}
\frac{\mathcal J^{AA}(T)}{\mathcal J^{AA}(0)}
&\simeq m_A(T)^{2-\varepsilon_{AA}},
\label{eq:eq_JAAm}\\
\frac{\mathcal J^{BB}(T)}{\mathcal J^{BB}(0)}
&\simeq m_B(T)^{2-\varepsilon_{BB}},
\label{eq:eq_JBBm}\\
\frac{\mathcal J^{AB}(T)}{\mathcal J^{AB}(0)}
&\simeq
\left[m_A(T)m_B(T)\right]^{1-\varepsilon_{AB}/2},
\label{eq:eq_JABm}
\end{align}
where $\varepsilon_{AA}$, $\varepsilon_{BB}$, and $\varepsilon_{AB}$ quantify the correlation-induced corrections to the corresponding mean-field exponents. For equivalent sublattices, $m_A=m_B=m$, all three expressions reduce to the conventional form $\mathcal J^{rs}(T)\propto m^{2-\varepsilon_{rs}}$\cite{Atxitia2010,Rzsa2023}. The uncorrelated mean-field limit is recovered for $\varepsilon_{rs}=0$, whereas positive values of $\varepsilon_{rs}$ describe a slower thermal suppression of the effective exchange interactions arising from transverse spin correlations.

Having established the quantitative agreement between the GF theory and ASD simulations for the representative ferrimagnetic system, we now use the GF formalism to investigate how these correlation corrections depend on the microscopic distribution of exchange interactions. To this end, we consider a family of rock-salt-type ferrimagnets parametrized by
\begin{align}
J_0^{AB}&=J_0^{BA}=-(1-\lambda)J_0,\label{eq:26a}
\\
J_0^{BB}&=\frac{1}{2}\lambda J_0,\label{eq:26b}
\\
J_0^{AA}&=\frac{3}{2}\lambda J_0,
\label{eq:26c}
\end{align}
Here, $\lambda\in[0,1]$ controls the partitioning between inter- and intrasublattice exchange while keeping the average zero-wave-vector intrasublattice exchange strength equal to $\lambda J_0$. At $\lambda=0$, the system is coupled exclusively through nearest-neighbor antiferromagnetic intersublattice exchange. Increasing $\lambda$ progressively transfers the exchange weight to the ferromagnetic intrasublattice interactions, while in the limit $\lambda\to1$ the intersublattice coupling vanishes and the system approaches two decoupled ferromagnetic sublattices.

The choice $J^{AA}_0=3J^{BB}_0$ explicitly breaks the equivalence between the two sublattices throughout the intermediate ferrimagnetic regime. To isolate the effect of this sublattice inequivalence, we also consider the corresponding equivalent-sublattice antiferromagnet,
\begin{equation}
J_0^{AA}=J_0^{BB}=\lambda J_0,
\qquad
J_0^{AB}=J_0^{BA}=-(1-\lambda)J_0,  
\label{eq:27}
\end{equation}
following Ref.~\cite{Rzsa2023}. For the ferrimagnetic system and for the numerical symmetric-AFM reference, the correlation corrections $\varepsilon_{rs}$ are extracted from low-temperature power-law fits to the GF results using Eqs.~\eqref{eq:eq_JAAm}--\eqref{eq:eq_JABm}. For the symmetric antiferromagnet, we additionally evaluate the analytical expressions derived in Appendix~\ref{appD}, with the Brillouin-zone averages
extrapolated to the thermodynamic limit. At the end points, the correlation corrections associated with exchange channels that vanish identically are understood as limiting values obtained as \(\lambda\) approaches the corresponding endpoint.

\begin{figure}
\centering
\includegraphics[width=1\linewidth]{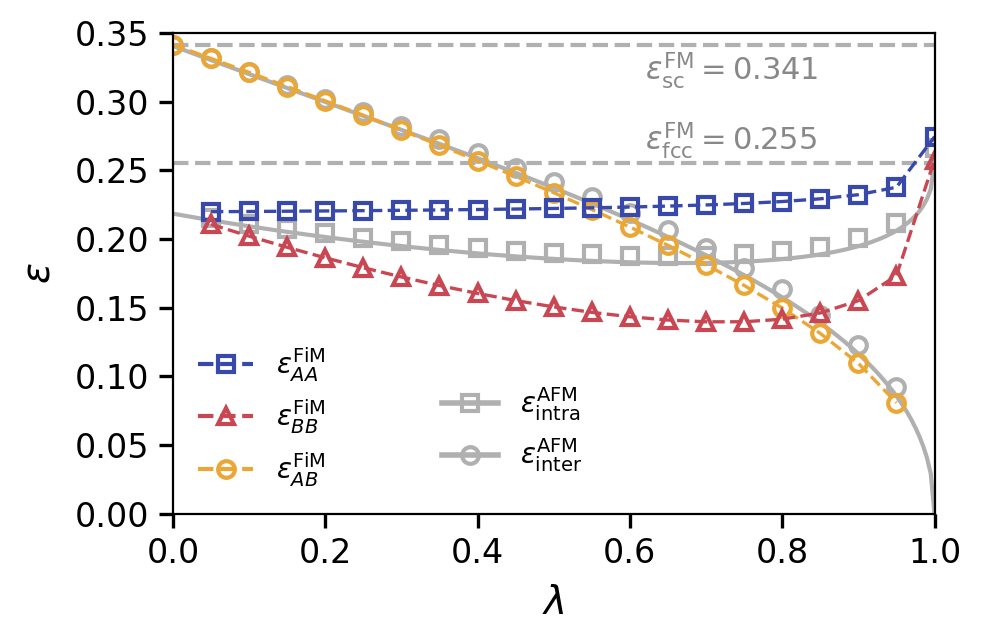}
\caption{Correlation corrections to the low-temperature exchange-scaling exponents in the rock-salt-type lattice as a function of the exchange-partition parameter $\lambda$. Colored symbols and lines correspond to the inequivalent-sublattice ferrimagnet defined by Eqs.~\eqref{eq:26a}--\eqref{eq:26c}, while gray symbols show the numerical GF results for the equivalent-sublattice antiferromagnet
defined by Eq.~\eqref{eq:27}. The solid gray curves correspond to the
analytical symmetric-AFM expressions derived in
Appendix~\ref{appD}, evaluated by extrapolating the discrete
Brillouin-zone sums to the thermodynamic limit.
The GF exponents are obtained from low-temperature power-law fits using
Eqs.~\eqref{eq:eq_JAAm}--\eqref{eq:eq_JABm}; their fitting
uncertainties are smaller than the symbol size. Horizontal dashed
lines indicate the simple-cubic and fcc ferromagnetic reference values,
$\varepsilon_{\mathrm{sc}}^{\mathrm{FM}}\simeq0.341$ and
$\varepsilon_{\mathrm{fcc}}^{\mathrm{FM}}\simeq0.255$, respectively.
Values associated with exchange channels that vanish at the endpoints
are understood as limiting values.}
\label{fig:scaling}
\end{figure}

Figure~\ref{fig:scaling} compares the correlation corrections for the
ferrimagnetic system with inequivalent magnetic sublattices, shown in
color, with those of the antiferromagnetic system with equivalent
sublattices, shown in gray. For the latter, the GF results are compared
with the analytical expressions derived in Appendix~\ref{appD}, whose
Brillouin-zone averages are extrapolated to the thermodynamic limit.
The numerical GF results closely follow the analytical reference
curves over the full range of $\lambda$. For more details on the procedure of extracting these corrections, see Appendix~\ref{appD}. In the antiferromagnetic system with equivalent sublattices, symmetry enforces a single intra-sublattice correction, $\varepsilon_{\mathrm{intra}}^{\mathrm{AFM}}$. Once this symmetry is broken in the ferrimagnet, the corresponding correction splits into two distinct quantities, $\varepsilon_{AA}^{\mathrm{FiM}}$ and $\varepsilon_{BB}^{\mathrm{FiM}}$. This splitting demonstrates that the thermal renormalization of an exchange channel is determined not only by its local exchange strength, \textcolor{black}{but also by the inter-sublattice coupling and by the distinct thermal evolution of the two sublattices.}
The evolution with $\lambda$ further illustrates this point. In the limit $\lambda\rightarrow0$, only the nearest-neighbor antiferromagnetic inter-sublattice exchange remains. For the bipartite rock-salt-type lattice considered here, a $\pi$ rotation of the spin direction on one sublattice maps this exchange network onto the corresponding nearest-neighbor ferromagnetic problem with simple-cubic connectivity. Consequently, 
the ferrimagnetic and antiferromagnetic inter-sublattice corrections therefore approach the same limiting value, corresponding to the simple-cubic ferromagnetic reference $\varepsilon_{\mathrm{sc}}^{\mathrm{FM}}\simeq0.341$~\cite{Rzsa2023}. In this regime, the details of the intra-sublattice exchange become irrelevant because the dominant thermal correlations are generated by the common inter-sublattice exchange network.
The opposite limit, $\lambda\rightarrow1$, provides a complementary picture. As the inter-sublattice coupling vanishes, the two sublattices become independent fcc ferromagnets. The intra-sublattice correlation corrections consequently converge toward the fcc ferromagnetic value $\varepsilon_{\mathrm{fcc}}^{\mathrm{FM}}\simeq0.255$, irrespective of the different absolute exchange strengths on the $A$ and $B$ sublattices. Thus, the splitting between $\varepsilon_{AA}^{\mathrm{FiM}}$ and $\varepsilon_{BB}^{\mathrm{FiM}}$ is not an intrinsic consequence of the unequal magnitudes of $J_0^{AA}$ and $J_0^{BB}$. Rather, it emerges from the coupled ferrimagnetic dynamics and disappears when the two sublattices become dynamically independent.
These results show that the scaling exponents cannot, in general, be regarded as universal constants determined solely by lattice geometry or by the type of the exchange interaction. In a ferrimagnet, they also depend on how the exchange is distributed among the intra- and inter-sublattice channels and on the resulting correlations between the two inequivalent magnetic subsystems. The parameter $\lambda$ therefore provides a direct means of interpolating between the correlation regimes characteristic of an inter-sublattice-dominated antiferromagnet and two independent ferromagnetic sublattices.

\section{Conclusions and outlook}

We have extended a 
GF framework to determine the temperature dependence of effective spin-model parameters in two-sublattice ferrimagnets with inequivalent and antiparallel magnetic sublattices. The approach provides the temperature-dependent intra- and inter-sublattice exchange interactions and magnetic anisotropies, together with the corresponding finite-temperature spin-wave spectrum. By comparison with 
ASD simulations, we have shown that the GF theory quantitatively reproduces the thermal evolution of the sublattice magnetizations, spin-wave excitations, and effective interaction parameters throughout most of the magnetically ordered phase.

A central result is that the different interaction channels exhibit distinct thermal renormalizations. The effective exchanges $\mathcal J^{AA}(T)$, $\mathcal J^{AB}(T)$, and $\mathcal J^{BB}(T)$ decrease at different rates because of the inequivalent thermal evolution and correlations of the two sublattices. The effective anisotropies show an even stronger thermal reduction and approximately follow the corresponding single-ion scaling with the sublattice magnetizations at low temperatures. The deviations of the exchange interactions from the uncorrelated scalings $m_A^2$, $m_A m_B$, and $m_B^2$ reflect transverse spin correlations that are naturally incorporated within the GF description.

We have further characterized these correlation effects through low-temperature scaling exponents. By continuously varying the relative strengths of the intra- and inter-sublattice exchange interactions, we find that the corresponding correlation corrections depend on the distribution of exchange among the different magnetic channels. In the coupled ferrimagnetic regime, breaking the equivalence between the two sublattices leads to distinct scaling exponents for the $AA$ and $BB$ exchange interactions. These differences disappear in the limit of two decoupled ferromagnetic sublattices, demonstrating that the scaling exponents are governed not only by lattice geometry and exchange type, but also by the correlations generated by the coupled ferrimagnetic environment.

The unequal thermal renormalization of the effective interactions has particularly important consequences for the long-wavelength dynamics. Although all exchange and anisotropy parameters decrease monotonically with temperature, the zone-center gaps and branch-resolved spin-wave curvature coefficients can display strongly nonmonotonic behavior. In particular, one circularly polarized branch develops an enhanced curvature in the vicinity of the angular-momentum-compensation region. Since these curvature coefficients characterize the spin-wave exchange stiffness and are directly connected to the exchange-stiffness parameters used in continuum descriptions, this result demonstrates that the effective long-wavelength dynamics cannot, in general, be inferred from the temperature dependence of the microscopic exchange interactions alone. Instead, they emerge from the combined temperature dependence of the exchange, anisotropy, sublattice magnetizations, and angular momenta.

These findings provide a direct route for transferring finite-temperature information from atomistic spin models to coarse-grained descriptions of ferrimagnetic dynamics. In particular, the GF approach offers a computationally efficient way to determine temperature-dependent micromagnetic parameters required as input for micromagnetic and multiscale models without relying exclusively on computationally demanding dynamical simulations at every temperature. This connection is especially relevant near compensation temperatures, where the strongly polarization- and temperature-dependent spin-wave stiffness may significantly influence the propagation and dynamics of magnetic textures and spin-wave excitations.

Several extensions of the present framework are of interest. Applying the method to material-specific Hamiltonians obtained from first-principles calculations would allow quantitative predictions for experimentally relevant ferrimagnets, including rare-earth--transition-metal alloys and compounds. The inclusion of additional interactions, such as anisotropic exchange, Dzyaloshinskii--Moriya interactions, and dipolar fields, would extend the description to more complex magnetic textures and nonreciprocal spin-wave dynamics. Furthermore, combining the temperature-dependent 
parameters derived here with a corresponding description of mode-dependent damping and lifetimes would enable predictions of spin-wave propagation lengths and dissipative dynamics. Finally, extending the present equilibrium treatment toward nonequilibrium situations, where the two sublattices may transiently acquire different 
degrees of magnetic order, would provide a natural route toward multiscale modeling of ultrafast ferrimagnetic dynamics.

Overall, the present work shows that the thermal evolution of ferrimagnetic dynamics is controlled not simply by a uniform softening of the underlying magnetic interactions, but by their unequal and correlation-dependent renormalization. Capturing this interplay is essential for a quantitative finite-temperature description of ferrimagnets and provides the microscopic basis for temperature-dependent micromagnetic and multiscale models.

\section*{ACKNOWLEDGMENTS}
M.Y. and U.A. acknowledge funding from PIE grant from CSIC (20226AT018), and the Spanish Ministerio de Ciencia, Innovación y Universidades grants (CNS2023-144681, PID2024-157112OB-C52, CEX2024-001445-S) funded by MCIN/AEI/10.13039/501100011033 and by ERDF A way of making Europe and ESF Investing in your future. 
J. M. L. and T. G. acknowledge financial support from PhD grants from Comunidad de Madrid (Grant No. PIPF-2023/TEC-29997 and No. PIPF-2022/TEC-25377). 
L.R. gratefully acknowledges funding by the National Research, Development, and Innovation Office (NRDI) of Hungary under Project Nos. FK142601 and ADVANCED 149745, as well as by the Hungarian Academy of Sciences via a János Bolyai Research Grant (No. BO/00178/23/11). 
R. M. O. and U. A. are grateful for financial support from COST Action CA23136 CHIROMAG.


\appendix

\section{Finite-temperature Green's-function theory}
\label{sec:app_A}

In this section, we present the explicit self-consistent Green's-function (GF) equations used to calculate the temperature-dependent sublattice magnetizations, spin correlations, and effective magnetic interactions. The formalism is based on the classical limit of the Callen--Anderson decoupling scheme and its extension to antiferromagnetically coupled two-sublattice systems.~\cite{Callen, Anderson, Rzsa2023}.

For the collinear ferrimagnetic state, it is convenient to work in local coordinate frames in which the equilibrium magnetization of both sublattices, $A$ and $B$, points along the positive local $z$ direction. We therefore define
\begin{equation}
    \widetilde{\mathbf S}_{iA}=\mathbf S_{iA},
    \qquad
    \widetilde S^{z}_{iB}=-S^{z}_{iB},
    \qquad
    \widetilde S^{\pm}_{iB}=-S^{\mp}_{iB}.
    \label{eq:SM_local_coordinates}
\end{equation}
With this convention, the reduced sublattice magnetizations are defined as
\begin{equation}
    m_r(T)\equiv
    \left\langle \widetilde S_r^z \right\rangle,
    \qquad r=A,B,
    \label{eq:SM_mr_definition}
\end{equation}
and remain positive throughout the ordered phase. The antiparallel orientation of the two physical sublattice magnetizations is recovered upon transforming back to the laboratory frame.
\onecolumngrid
\subsection{Temperature-dependent dynamical matrix}
\label{subsec:SM_GF_matrix}
Linearizing the equation of motion introduced in the main text around the collinear ferrimagnetic equilibrium state and retaining terms to first order in the transverse spin components yields a coupled spin-wave equation for the two sublattices. In the local coordinate system defined above, we define
\begin{equation}
\widetilde{\mathbf S}^{(1)}_{\bm q}
=
\begin{pmatrix}
    \widetilde S^{+}_{\bm q A}\\[1mm]
    \widetilde S^{-}_{\bm q B}
\end{pmatrix},
\qquad
\widetilde{\mathbf S}^{(2)}_{\bm q}
=
\begin{pmatrix}
    \widetilde S^{-}_{\bm q A}\\[1mm]
    \widetilde S^{+}_{\bm q B}
\end{pmatrix}.
\label{eq:SM_transverse_vectors}
\end{equation}
The two sectors are related by complex conjugation and yield the same
spin-wave spectrum. In the following, we write explicitly the equation
for $\widetilde{\mathbf S}^{(1)}_{\bm q}$.

After spatial and temporal Fourier transformation, the equation of motion takes the form

\begin{equation}
\omega
\begin{pmatrix}
    \widetilde S^{+}_{\bm q A}\\[1mm]
    \widetilde S^{-}_{\bm q B}
\end{pmatrix}
=
\begin{pmatrix}
    \dfrac{\gamma_A}{\mu_A} & 0\\[2mm]
    0 & \dfrac{\gamma_B}{\mu_B}
\end{pmatrix}
\underbrace{
\begin{pmatrix}
     J^{AA}_{\bm 0}
    +2K_A
    - J^{AB}_{\bm 0}
    - J^{AA}_{\bm q}
    &
    J^{AB}_{\bm q}
    \\[2mm]
    - J^{BA}_{\bm q}
    &
    - J^{BB}_{\bm 0}
    -2K_B
    + J^{BA}_{\bm 0}
    + J^{BB}_{\bm q}
\end{pmatrix}
}_{\displaystyle \widetilde{H}_{\mathrm{SW},\bm q}}
\begin{pmatrix}
    \widetilde S^{+}_{\bm q A}\\[1mm]
    \widetilde S^{-}_{\bm q B}
\end{pmatrix}.
\label{eq:SM_linearized_eom}
\end{equation}

Here $\widetilde{H}_{\textrm{SW}, \bm{q}}$ is the 
spin-wave matrix in the harmonic approximation and 
\begin{equation}
     J^{rs}_{\bm q}
    =
    \sum_j
    J^{rs}_{ij}
    e^{-i\bm q\cdot\mathbf R_{ij}},
    \qquad
    \mathbf R_{ij}=\mathbf R_i-\mathbf R_j,
    \label{eq:SM_Fourier_exchange}
\end{equation}
is the Fourier transform of the atomistic exchange interactions. Owing to translational invariance, $ J^{rs}_{\bm q}$ is independent of the choice of the reference site $i$. The quantities $J^{rs}_{ij}$ and $K_r$ are the temperature-independent parameters of the atomistic Hamiltonian.

At finite temperature, the thermal population of the magnon modes reduces the sublattice magnetizations and generates transverse spin correlations.  Within the Green's-function formalism, this feedback is incorporated self-consistently, and the single-particle excitation spectrum is given by the eigenvalues of
$\text{diag}\left(\gamma_A/\mu_A, \gamma_B/\mu_B\right)\widetilde{\boldsymbol{\Gamma}}_{\bm q}$, where the temperature-dependent matrix $\widetilde{\Gamma}_{\bm{q}}$ replaces the spin-wave Hamiltonian $\widetilde{H}_{SW, \bm{q}}$ in Eq. \eqref{eq:SM_linearized_eom}, with elements

\begin{align}
\widetilde{\Gamma}^{AA}_{\bm q}
={}&
 J^{AA}_{\bm 0}m_A
+2K_A m_A
- J^{AB}_{\bm 0}m_B
- J^{AA}_{\bm q}m_A
-2\alpha_0
\sum_{\bm q'}
\Big[
\big(
 J^{AA}_{\bm q-\bm q'}
- J^{AA}_{\bm q'}
\big)
\Phi^{AA}_{\bm q'}
m^{2}_A
+
 J^{AB}_{\bm q'}
\Phi^{AB}_{\bm q'}
m_A m_B
\Big],
\label{eq:SM_Gamma_AA}
\\[2mm]
\widetilde{\Gamma}^{AB}_{\bm q}
={}&
 J^{AB}_{\bm q}m_A
+
2\alpha_0
\sum_{\bm q'}
 J^{AB}_{\bm q-\bm q'}
\Phi^{BA}_{\bm q'}m_A m_B,
\label{eq:SM_Gamma_AB}
\\[2mm]
\widetilde{\Gamma}^{BA}_{\bm q}
={}&
- J^{BA}_{\bm q}m_B
-
2\alpha_0
\sum_{\bm q'}
 J^{BA}_{\bm q-\bm q'}
\Phi^{AB}_{\bm q'}m_A m_B,
\label{eq:SM_Gamma_BA}
\\[2mm]
\widetilde{\Gamma}^{BB}_{\bm q}
={}&
- J^{BB}_{\bm 0}m_B
-2K_B m_B
+ J^{BA}_{\bm 0}m_A
+ J^{BB}_{\bm q}m_B
+
2\alpha_0
\sum_{\bm q'}
\Big[
\big(
 J^{BB}_{\bm q-\bm q'}
- J^{BB}_{\bm q'}
\big)
\Phi^{BB}_{\bm q'}
m^{2}_B
+
 J^{BA}_{\bm q'}
\Phi^{BA}_{\bm q'}
m_A m_B
\Big].
\label{eq:SM_Gamma_BB}
\end{align}

The microscopic parameters $J^{rs}_{ij}$ and $K_r$ entering
$\widetilde{\boldsymbol{\Gamma}}_{\bm q}$ remain those of the
temperature-independent atomistic Hamiltonian; its temperature
dependence arises through $m_r$ and $\Phi_{\bm q}^{rs}$. Here, $\Phi^{rs}_{\bm q}$ denotes the transverse spin-correlation functions, defined below, while the coefficient $\alpha_0$ specifies the decoupling of the higher-order Green's functions. Throughout this work, we set $\alpha_0=1/2$, corresponding to the classical limit of the Callen--Anderson decoupling scheme \cite{Callen,Anderson,Rzsa2023}.

The eigenvalues of the temperature-dependent dynamical matrix yield the two magnon branches of the ferrimagnet,

\begin{equation}
    \omega_{\bm{q}}^{\pm} = \frac{1}{2} \left(\frac{\gamma_A}{\mu_A}\widetilde{\Gamma}^{AA}_{\bm q} + \frac{\gamma_B}{\mu_B}\widetilde{\Gamma}^{BB}_{\bm q} \right) \pm \frac{1}{2} \sqrt{\left(\frac{\gamma_A}{\mu_A}\widetilde{\Gamma}^{AA}_{\bm q} - \frac{\gamma_B}{\mu_B}\widetilde{\Gamma}^{BB}_{\bm q} \right)^2 + 4 \frac{\gamma_A}{\mu_A}\widetilde{\Gamma}^{AB}_{\bm q} \frac{\gamma_B}{\mu_B}\widetilde{\Gamma}^{BA}_{\bm q} }
    \label{eq:SM_magnon_frequencies}
\end{equation}


\twocolumngrid

\subsection{Spin correlations and sublattice magnetizations}
\label{subsec:SM_GF_correlations}

The matrix $\boldsymbol{\Phi}_{\bm q}$ entering Eqs.~\eqref{eq:SM_Gamma_AA}--\eqref{eq:SM_Gamma_BB}
is obtained from the spectral theorem. In the classical limit, where the magnon occupation is $n(\omega)=k_{\mathrm B}T/\omega$, it reduces to
\begin{equation}
    \boldsymbol{\Phi}_{\bm q}
    =
    \frac{k_{\mathrm B}T}{N_{\mathrm c}}
    \boldsymbol{\sigma}^{z}
    \left(
    \widetilde{\boldsymbol{\Gamma}}_{\bm q}^{-1}
    \right)^{T},
    \qquad
    \boldsymbol{\sigma}^{z}
    =
    \begin{pmatrix}
    1 & 0\\
    0 & -1
    \end{pmatrix},
    \label{eq:SM_Phi_matrix}
\end{equation}
where $N_{\mathrm c}$ is the number of unit cells. Explicitly,
\begin{align}
    \Phi^{AA}_{\bm q}
    &=
    \frac{k_{\mathrm B}T}{N_{\mathrm c}}
    \frac{\widetilde{\Gamma}^{BB}_{\bm q}}
    {\det\widetilde{\boldsymbol{\Gamma}}_{\bm q}},
    \label{eq:SM_Phi_AA}
    \\
    \Phi^{AB}_{\bm q}
    &=
    -\frac{k_{\mathrm B}T}{N_{\mathrm c}}
    \frac{\widetilde{\Gamma}^{BA}_{\bm q}}
    {\det\widetilde{\boldsymbol{\Gamma}}_{\bm q}},
    \label{eq:SM_Phi_AB}
    \\
    \Phi^{BA}_{\bm q}
    &=
    \frac{k_{\mathrm B}T}{N_{\mathrm c}}
    \frac{\widetilde{\Gamma}^{AB}_{\bm q}}
    {\det\widetilde{\boldsymbol{\Gamma}}_{\bm q}},
    \label{eq:SM_Phi_BA}
    \\
    \Phi^{BB}_{\bm q}
    &=
    -\frac{k_{\mathrm B}T}{N_{\mathrm c}}
    \frac{\widetilde{\Gamma}^{AA}_{\bm q}}
    {\det\widetilde{\boldsymbol{\Gamma}}_{\bm q}},
    \label{eq:SM_Phi_BB}
\end{align}
with
\begin{equation}
    \det\widetilde{\boldsymbol{\Gamma}}_{\bm q}
    =
    \widetilde{\Gamma}^{AA}_{\bm q}
    \widetilde{\Gamma}^{BB}_{\bm q}
    -
    \widetilde{\Gamma}^{AB}_{\bm q}
    \widetilde{\Gamma}^{BA}_{\bm q}.
    \label{eq:SM_detGamma}
\end{equation}

The matrix $\boldsymbol{\Phi}_{\bm q}$ is directly related to the
transverse spin-correlation functions introduced above. With the
definitions in Eq.~\eqref{eq:SM_transverse_vectors},
\begin{equation}
\left\langle
\widetilde{\mathbf S}^{(1)}_{-\bm q}
\left(
\widetilde{\mathbf S}^{(2)}_{\bm q}
\right)^{T}
\right\rangle
=
2
\begin{pmatrix}
m_A & 0\\
0 & m_B
\end{pmatrix}
\boldsymbol{\Phi}_{\bm q}.
\label{eq:SM_transverse_correlations}
\end{equation}
The real-space transverse correlations entering the effective
interaction parameters below are obtained by inverse Fourier
transformation of Eq.~\eqref{eq:SM_transverse_correlations}.

The sublattice magnetizations follow self-consistently from the classical Langevin relation
\begin{equation}
    m_r(T)
    =
    \coth\left(\frac{1}{\Phi_r}\right)
    -
    \Phi_r,
    \qquad
    \Phi_r
    =
    \sum_{\bm q}
    \Phi^{rr}_{\bm q}.
    \label{eq:SM_Langevin}
\end{equation}
At zero temperature $\Phi_r\rightarrow0$, 
therefore $m_r\rightarrow1$.

The main result of the GF formalism is the calculation of the frequencies of the two magnon modes $\omega_{\bm{q}}^{\pm}$  in Eq. \eqref{eq:SM_magnon_frequencies} and of the sublattice magnetization in Eq. \eqref{eq:SM_Langevin}.

\subsection{Effective exchange and anisotropy parameters}
\label{subsec:SM_GF_effective}

The connection between the microscopic spin model and its long-wavelength description can be established by requiring that both models yield the same spin-wave spectrum as $\bm q\rightarrow0$. At the continuum level, we write the magnetization of sublattice $r$ as
\begin{equation}
    \mathbf M_r(\mathbf r,T)
    =
    M_r(T)\mathbf n_r(\mathbf r),
    \qquad
    |\mathbf n_r|=1,
    \label{eq:SM_continuum_magnetization}
\end{equation}
where $M_r(T)=\mu_r m_r(T)/V_{\mathrm c}$ denotes the magnitude of the sublattice magnetization density in the local frame.

For the isotropic exchange and single-ion anisotropy considered here, the long-wavelength free-energy functional can be written as
\begin{equation}
\begin{aligned}
    \mathcal F
    ={}&
    \int d^d\mathbf r
    \Bigg[
    \frac{1}{2}
    \sum_{r,s}
    \sum_{\alpha,\beta,\gamma}
    \mathscr J_m^{rs,\alpha\beta}
    \,
    \partial_\alpha n_r^\gamma
    \partial_\beta n_s^\gamma
    \\&\quad\quad\quad\quad-
    \sum_r
    \mathscr K_m^r
    (n_r^z)^2
    -
    \mathscr J_{m0}^{AB}
    \mathbf n_A\cdot\mathbf n_B
    \Bigg].
    \label{eq:SM_free_energy}
\end{aligned}
\end{equation}
Here $\mathscr J_m^{rs,\alpha\beta}$ describes the exchange cost
associated with spatially nonuniform configurations,
$\mathscr K_m^r$ is the effective single-ion anisotropy of sublattice
$r$, and $\mathscr J_{m0}^{AB}$ is the homogeneous intersublattice
exchange contribution.

The equation of motion associated with
Eq.~\eqref{eq:SM_free_energy} reads
\begin{equation}
    \partial_t\mathbf n_r
    =
    -\gamma_r
    \mathbf n_r
    \times
    \left(
    -\frac{1}{M_r(T)}
    \frac{\delta\mathcal F}{\delta\mathbf n_r}
    \right).
    \label{eq:SM_continuum_eom}
\end{equation}
Matching the continuum excitation spectrum to the long-wavelength
limit of the microscopic GF spectrum gives the following temperature-dependent continuum coefficients:
\begin{align}
\mathscr J_m^{rs,\alpha\beta}(T)
={}&
\frac{1}{2V_{\mathrm c}}
\sum_j
\left[
J_{ij}^{rs}
+
\alpha_0J_{ij}^{rs}
\operatorname{Re}
\left\langle
\widetilde S_{js}^{(1)}
\widetilde S_{ir}^{(2)}
\right\rangle
\right]
\notag\\
&\quad\quad\times
R_{ij}^{\alpha}
R_{ij}^{\beta}
m_r(T)m_s(T),
\label{eq:SM_Jm}
\\[2mm]
\mathscr K_m^{r}(T)
={}&
\frac{K_r}{V_{\mathrm c}}
\left[
1-
\alpha_0
\left\langle
\widetilde S_{ir}^{(1)}
\widetilde S_{ir}^{(2)}
\right\rangle
\right]
m_r^2(T),
\label{eq:SM_Km}
\\[2mm]
\mathscr J_{m0}^{AB}(T)
={}&
\frac{1}{V_{\mathrm c}}
\sum_j
\left[
J_{ij}^{AB}
+
\alpha_0J_{ij}^{AB}
\operatorname{Re}
\left\langle
\widetilde S_{jB}^{-}
\widetilde S_{iA}^{-}
\right\rangle
\right]
\notag\\
&\quad\quad\times
m_A(T)m_B(T).
\label{eq:SM_Jm0}
\end{align}

Equations~\eqref{eq:SM_Jm}--\eqref{eq:SM_Jm0} separate two
contributions to the thermal renormalization. The factors
$m_r m_s$ constitute the uncorrelated, or random-phase, contribution,
whereas the terms proportional to $\alpha_0$ account explicitly for
transverse spin correlations. Setting $\alpha_0=0$ therefore recovers
the random-phase result. For $\alpha_0=1/2$, transverse correlations
reduce the thermal suppression of the exchange-related coefficients, whereas they enhance the thermal reduction of the anisotropy coefficient. This distinction underlies the
different magnetization scalings of exchange and anisotropy discussed
in the main text.

\subsection{Self-consistent numerical solution}
\label{subsec:SM_GF_numerics}

Equations~\eqref{eq:SM_Gamma_AA}--\eqref{eq:SM_Langevin} form a
closed set of nonlinear self-consistency equations. At $T=0$ the
calculation is initialized with
\begin{equation}
    m_A=m_B=1,
    \qquad
    \boldsymbol{\Phi}_{\bm q}=0.
    \label{eq:SM_initial_condition}
\end{equation}
For each finite temperature, the converged solution at the preceding
temperature is used as the initial condition. Starting from the current
estimate of
$\widetilde{\boldsymbol{\Gamma}}_{\bm q}$,
Eq.~\eqref{eq:SM_Phi_matrix} is evaluated  on a uniform reciprocal-space
grid spanning the first Brillouin zone. The diagonal components are then summed over
$\bm q$ to obtain $\Phi_A$ and $\Phi_B$, and the sublattice
magnetizations are updated using Eq.~\eqref{eq:SM_Langevin}.
The transverse correlation functions, 
 together with the updated sublattice magnetizations are then inserted into
Eqs.~\eqref{eq:SM_Gamma_AA}--\eqref{eq:SM_Gamma_BB} to obtain an
updated dynamical matrix.

This procedure is iterated until both sublattice magnetizations and all
matrix elements
$\widetilde{\Gamma}^{rs}_{\bm q}$
are converged within the prescribed numerical tolerance over the full
reciprocal-space mesh. Once self-consistency is reached, the resulting
matrix determines the finite-temperature magnon spectrum through Eq.~\eqref{eq:SM_magnon_frequencies}, while
Eqs.~\eqref{eq:SM_Jm}--\eqref{eq:SM_Jm0} provide the corresponding
effective interaction parameters. Repeating this procedure as a
function of temperature yields the GF curves reported in the main
text.

\onecolumngrid
\section{Long-wavelength expansion of the spin-wave spectrum}
\label{sec:appB}


In this section, we derive the long-wavelength form of the two magnon
branches in terms of effective temperature-dependent atomistic
parameters. We denote by $\mathcal J^{rs}(T)$ and $\mathcal K_r(T)$
the thermally renormalized exchange interaction per bond and the
thermally renormalized on-site anisotropy, respectively. These
quantities have the same units and lattice-level meaning as the bare
microscopic parameters $J_{ij}^{rs}$ and $K_r$ entering the atomistic
Hamiltonian, but include the thermal renormalization arising from the
sublattice magnetizations and transverse spin correlations.

For the exchange geometry considered here, all equivalent bonds within
a given interaction shell share the same effective exchange parameter
$\mathcal J^{rs}(T)$. Its lattice Fourier transform is therefore
defined as
\begin{equation}
    \mathcal J_{\bm q}^{rs}(T)
    =
    \mathcal J^{rs}(T)
    \sum_{\bm R_{ij}\in rs}
    e^{-i\bm q\cdot\bm R_{ij}}.
    \label{eq:SM_effective_Jq}
\end{equation}
The corresponding longitudinal sublattice moments are
$\mu_r m_r(T)$.

The matrix $\boldsymbol{\Gamma}_{\bm q}$ introduced below denotes the
interaction matrix of this effective atomistic spin model and should
not be confused with the self-consistent Green's-function matrix
$\widetilde{\boldsymbol{\Gamma}}_{\bm q}$ defined in Appendix~A.
In the latter, the magnetization and correlation corrections appear
explicitly, whereas here they are absorbed into the effective
parameters $\mathcal J^{rs}(T)$ and $\mathcal K_r(T)$.

\begin{equation}
\omega_{\bm q}^{\pm}(T)
=
\frac{1}{2}
\left(
\frac{\gamma_A}{\mu_A m_A}\Gamma^{AA}_{\bm q}
+
\frac{\gamma_B}{\mu_B m_B}\Gamma^{BB}_{\bm q}
\right)
\pm
\frac{1}{2}
\sqrt{
\left(
\frac{\gamma_A}{\mu_A m_A}\Gamma^{AA}_{\bm q}
-
\frac{\gamma_B}{\mu_B m_B}\Gamma^{BB}_{\bm q}
\right)^2
+
4
\frac{\gamma_A}{\mu_A m_A}\Gamma^{AB}_{\bm q}
\frac{\gamma_B}{\mu_B m_B}\Gamma^{BA}_{\bm q}
}.
\label{eq:SM_dispersion}
\end{equation}

The same expression applies at any fixed temperature upon replacing
the bare interactions by their effective temperature-dependent
counterparts. 

For the exchange geometry considered here, the interaction coefficients
entering Eq.~\eqref{eq:SM_dispersion} are

\begin{align}
\Gamma^{AA}_{\bm q}
&=
\mathcal J^{AA}_{\bm 0}(T)
-
\mathcal J^{AA}_{\bm q}(T)
-
\mathcal J^{AB}_{\bm 0}(T)
+
2\mathcal K_A(T),\label{eq:SM_Gamma_lw_a}
\qquad
&\\
\Gamma^{AB}_{\bm q}
&=
\mathcal J^{AB}_{\bm q}(T),\label{eq:SM_Gamma_lw_b}
\\
\Gamma^{BB}_{\bm q}
&=
-
\mathcal J^{BB}_{\bm 0}(T)
+
\mathcal J^{BB}_{\bm q}(T)
+
\mathcal J^{BA}_{\bm 0}(T)
-
2\mathcal K_B(T),
\qquad\label{eq:SM_Gamma_lw_c}
&\\
\Gamma^{BA}_{\bm q}
&=
-
\mathcal J^{BA}_{\bm q}(T).
\label{eq:SM_Gamma_lw_d}
\end{align}

For equivalent $AB$ and $BA$ bonds,
$\mathcal J^{AB}_{\bm q}(T)=\mathcal J^{BA}_{\bm q}(T)$. In the following, the explicit temperature dependence of
$\mathcal J^{rs}(T)$, $\mathcal J_{\bm q}^{rs}(T)$,
$\mathcal K_r(T)$, and $m_r(T)$ is omitted for compactness.

\subsection{Expansion around the Brillouin-zone center}
\label{subsec:SM_lw_expansion}

To obtain the long-wavelength dispersion, we expand the Fourier
transforms of the exchange interactions around the Brillouin-zone
center, $\bm q=\bm 0$. For the rock-salt lattice, the $AA$ and $BB$
intrasublattice interactions connect twelve second-nearest neighbors,
whereas the $AB$ interaction connects six nearest neighbors.

The corresponding long-wavelength expansions of the effective
exchange Fourier transforms are
\begin{align}
\mathcal J_{\bm q}^{AA}
&=
12\mathcal J^{AA}
-
4\mathcal J^{AA}(qa)^2
+
\mathcal O[(qa)^4], \label{eq:SM_JAA_expansion}
\\
\mathcal J_{\bm q}^{AB}
&=
6\mathcal J^{AB}
-
\mathcal J^{AB}(qa)^2
+
\mathcal O[(qa)^4], \label{eq:SM_JAB_expansion}
\\
\mathcal J_{\bm q}^{BB}
&=
12\mathcal J^{BB}
-
4\mathcal J^{BB}(qa)^2
+
\mathcal O[(qa)^4].\label{eq:SM_JBB_expansion}
\end{align}

where $q^2=q_x^2+q_y^2+q_z^2$.  Thus, to quadratic order in $\bm q$, cubic symmetry makes the
long-wavelength dispersion isotropic.
Substituting
Eqs.~\eqref{eq:SM_JAA_expansion}--\eqref{eq:SM_JBB_expansion}
into Eqs.~\eqref{eq:SM_Gamma_lw_a}--\eqref{eq:SM_Gamma_lw_d} gives
\begin{align}
\Gamma^{AA}_{\bm q}
&\simeq
-6\mathcal{J}^{AB}
+
2 \mathcal K_A
+
4\mathcal{J}^{AA}(qa)^2,
\label{eq:SM_GammaAA_exp}
\\
\Gamma^{AB}_{\bm q}
&\simeq
6\mathcal{J}^{AB}
-
\mathcal{J}^{AB}(qa)^2,
\label{eq:SM_GammaAB_exp}
\\
\Gamma^{BB}_{\bm q}
&\simeq
6\mathcal{J}^{AB}
-
2 \mathcal K_B
-
4\mathcal{J}^{BB}(qa)^2,
\label{eq:SM_GammaBB_exp}
\\
\Gamma^{BA}_{\bm q}
&\simeq
-6\mathcal{J}^{AB}
+
\mathcal{J}^{AB}(qa)^2.
\label{eq:SM_GammaBA_exp}
\end{align}
The first term in Eq.~\eqref{eq:SM_dispersion} can then be expanded as
\begin{equation}
\begin{aligned}
\frac{\gamma_A}{\mu_A m_A}\Gamma^{AA}_{\bm q}
+
\frac{\gamma_B}{\mu_B m_B}\Gamma^{BB}_{\bm q}
={}&
2\frac{\gamma_A}{\mu_A m_A}\mathcal K_A
-
2\frac{\gamma_B}{\mu_B m_B}\mathcal K_B
\\
&-
6\mathcal J^{AB}
\left(
\frac{\gamma_A}{\mu_A m_A}
-
\frac{\gamma_B}{\mu_B m_B}
\right)
+
4\left(
\frac{\gamma_A}{\mu_A m_A}\mathcal J^{AA}
-
\frac{\gamma_B}{\mu_B m_B}\mathcal J^{BB}
\right)(qa)^2.
\end{aligned}
\label{eq:SM_trace_expansion}
\end{equation}
Likewise, the argument of the square root in
Eq.~\eqref{eq:SM_dispersion} can be written as
\begin{equation}
\left(
\frac{\gamma_A}{\mu_A m_A}\Gamma^{AA}_{\bm q}
-
\frac{\gamma_B}{\mu_B m_B}\Gamma^{BB}_{\bm q}
\right)^2
+
4\frac{\gamma_A}{\mu_A m_A}\frac{\gamma_B}{\mu_B m_B}
\Gamma^{AB}_{\bm q}\Gamma^{BA}_{\bm q}
=
D_0
+
D_2(qa)^2
+
\mathcal O\!\left[(qa)^4\right].
\label{eq:SM_discriminant_expansion}
\end{equation}
where we have introduced the auxiliary coefficients,
\begin{align}
D_0
&=
\left(
2\frac{\gamma_A}{\mu_A m_A}\mathcal K_A
+
2\frac{\gamma_B}{\mu_B m_B} \mathcal K_B
-
6\mathcal{J}^{AB}\left(\frac{\gamma_A}{\mu_A m_A}+\frac{\gamma_B}{\mu_B m_B}\right)\right)^2
-
4\frac{\gamma_A}{\mu_A m_A}\frac{\gamma_B}{\mu_B m_B}(6\mathcal{J}^{AB})^2,
\label{eq:SM_D0}
\\
D_2
&=
8\left(
2\frac{\gamma_A}{\mu_A m_A}\mathcal K_A
+
2\frac{\gamma_B}{\mu_B m_B}\mathcal K_B
-
6\mathcal{J}^{AB}\left(\frac{\gamma_A}{\mu_A m_A}+\frac{\gamma_B}{\mu_B m_B}\right)\right)\left(
\frac{\gamma_A}{\mu_A m_A}\mathcal{J}^{AA}
+
\frac{\gamma_B}{\mu_B m_B}\mathcal{J}^{BB}
\right)
\\ 
&\quad+
48\frac{\gamma_A}{\mu_A m_A}\frac{\gamma_B}{\mu_B m_B}(\mathcal{J}^{AB})^2.
\label{eq:SM_D2}
\end{align}
The square-root contribution therefore becomes
\begin{equation}
\sqrt{
D_0+D_2(qa)^2
}
\simeq
\sqrt{D_0}
+
\frac{D_2}{2\sqrt{D_0}}(qa)^2
+
\mathcal O\!\left[(qa)^4\right].
\label{eq:SM_sqrt_expansion}
\end{equation}

\subsection{Zone-center spin-wave gaps}
\label{subsec:SM_gaps}

Combining Eqs.~\eqref{eq:SM_trace_expansion} and
\eqref{eq:SM_sqrt_expansion}, the two magnon branches take the
long-wavelength form
\begin{equation}
\omega_{\bm q}^{\pm}
=
\omega_0^{\pm}
+
\mathcal A^{\pm}(qa)^2
+
\mathcal O\!\left[(qa)^4\right],
\label{eq:SM_long_wavelength}
\end{equation}
where $\omega_0^{\pm}$ are the zone-center frequencies. They are given
by
%
%
\begin{equation}
\begin{aligned}
\omega_0^\pm
={}&
\frac{1}{2}
\Bigg[
2\frac{\gamma_A}{\mu_A m_A}\mathcal{K_A}
-
2\frac{\gamma_B}{\mu_B m_B}\mathcal{K_B}
-
6\mathcal{J^{AB}}
\left(
\frac{\gamma_A}{\mu_A m_A}
-
\frac{\gamma_B}{\mu_B m_B}
\right)
\Bigg]
\pm
\frac{1}{2}\sqrt{D_0}.
\end{aligned}
\label{eq:SM_omega0}
\end{equation}
Equation~\eqref{eq:SM_omega0} shows that the zone-center
frequencies are determined by the intersublattice exchange and the
two anisotropy contributions. In contrast, the intrasublattice
exchange interactions $\mathcal{J}^{AA}$ and $\mathcal{J}^{BB}$ do not contribute at
$\bm q=\bm 0$, since a spatially uniform precession does not generate
relative rotations between spins belonging to the same sublattice.\\


\twocolumngrid

\subsection{Branch-resolved spin-wave stiffness coefficients}
\label{subsec:SM_stiffness}
The quadratic dispersion coefficients in
Eq.~\eqref{eq:SM_long_wavelength} can be written
explicitly as
\begin{equation}
\mathcal A^{\pm}
=
2\left(
\frac{\gamma_A}{\mu_A m_A}\mathcal{J}^{AA}
-
\frac{\gamma_B}{\mu_B m_B}\mathcal{J}^{BB}
\right)
\pm
\frac{
D_2
}
{4\sqrt{D_0}}.
\label{eq:SM_stiffness_explicit}
\end{equation}

\textcolor{black}{Equation~\eqref{eq:SM_stiffness_explicit} assumes a finite separation of the two
zone-center modes, $D_0>0$, and is valid in the regime
$D_2(qa)^2\ll D_0$. As $D_0$ approaches zero, the range of wave vectors
over which the quadratic expansion is valid becomes progressively
smaller, and the apparent $1/\sqrt{D_0}$ enhancement of $A^\pm$ should
not be interpreted as a physical divergence. At $D_0=0$, the square-root
term must instead be retained in its unexpanded form; for $D_2\neq0$ it
behaves as $\sqrt{D_2}\,|qa|$, so that the dispersion is no longer
quadratic and the coefficients $A^\pm$ are not well defined.}
Unlike the zone-center frequencies, the curvature
coefficients depend on all three exchange channels.
Their temperature dependence is therefore determined
by the combined renormalization of the intra- and
intersublattice exchange interactions and the
sublattice magnetizations, rather than by any
individual exchange parameter.

Consequently, the temperature dependence of
$\mathcal A^{\pm}(T)$ reflects the competition between the different
exchange channels and can be nonmonotonic even when each individual
effective exchange interaction decreases monotonically with
temperature.

\section{Extraction of effective exchange and anisotropy parameters}
\label{appC:parameter_extraction}

The effective interaction parameters associated with the
atomistic spin-dynamics (ASD) spectra are extracted by exploiting the
different information carried by the dispersive and zone-center parts
of the magnon spectrum. As shown in Appendix~\ref{sec:appB}, the
intrasublattice exchanges $\mathcal J^{AA}$ and
$\mathcal J^{BB}$ do not contribute to the zone-center frequencies,
whereas the dispersion away from $\Gamma$ is strongly controlled by
the exchange interactions. This naturally separates the information
required to determine the exchange and anisotropy parameters.

A simultaneous five-parameter fit of
$\mathcal J^{AB}$, $\mathcal J^{AA}$,
$\mathcal J^{BB}$, $\mathcal K_A$, and $\mathcal K_B$
is poorly conditioned for the anisotropies. Although the ASD spectrum
contains a large number of reciprocal-space points, most of these
sample the exchange-dominated dispersive region and therefore provide
little independent information on $\mathcal K_A$ and $\mathcal K_B$.
Small changes in the anisotropies can consequently produce only minor
changes in the global residual and can be partially compensated by
variations of $\mathcal J^{AB}$. We therefore determine the exchange
interactions from the finite-$q$ spectrum and obtain the anisotropies
separately from the two zone-center frequencies.

\subsection{Determination of the exchange parameters}
\label{subsec:SM_J_extraction}

At each temperature, the three effective exchange parameters
$\mathcal J^{AB}$, $\mathcal J^{AA}$, and
$\mathcal J^{BB}$ are obtained from a simultaneous nonlinear
least-squares fit to both ASD magnon branches. During this step,
$\mathcal K_A$ and $\mathcal K_B$ are kept fixed. To minimize the influence
of the anisotropy-induced zone-center gaps, only wave vectors
satisfying
\begin{equation}
    |\bm q| \ge q_J^{\mathrm{min}},
    \qquad
    q_J^{\mathrm{min}}=0.15,
    \label{eq:SM_J_qmin}
\end{equation}
are included in the fit. Here and in the following,
$\bm q$ denotes the reduced wave vector expressed in units of
$\pi/a$.

Because the exchange interactions determine the dispersion over a
large fraction of the Brillouin zone, the resulting
$\mathcal J^{AB}$, $\mathcal J^{AA}$, and
$\mathcal J^{BB}$ are well constrained and were found to be largely
insensitive to moderate variations of the fitting procedure.

\subsection{Extraction of the anisotropy parameters and uncertainty estimation}
\label{subsec:SM_K_extraction}

The anisotropy parameters are determined from the zone-center
frequencies of the two magnon branches. As shown in
Appendix~\ref{sec:appB}, the intrasublattice exchange
contributions $\mathcal J^{AA}$ and $\mathcal J^{BB}$ cancel
exactly at $\bm q=0$. Consequently, the two frequencies
$\omega_0^{+}$ and $\omega_0^{-}$ depend only on
$\mathcal J^{AB}$, $\mathcal K_A$, and $\mathcal K_B$,
through Eq.~\eqref{eq:SM_omega0}.

At each temperature, $\mathcal J^{AB}$ is fixed to the value obtained
from the finite-$q$ fit described above, while $m_A(T)$ and $m_B(T)$
are taken directly from the ASD simulations. The two zone-center
frequencies therefore provide two equations for the two remaining
unknowns, $\mathcal K_A$ and $\mathcal K_B$, which are determined
simultaneously without introducing any additional fitting parameters.

The values of $\omega_0^\pm$ are obtained from the ASD spectrum in the
vicinity of $\Gamma$. We verified that using the directly simulated
$\Gamma$-point frequencies or a local extrapolation from neighboring
wave vectors leads to indistinguishable anisotropy parameters within
the accuracy relevant here.

In contrast to the exchange parameters, which are constrained by a
large portion of the dispersive spectrum, the anisotropies are
determined from only the two zone-center frequencies and the value of
$\mathcal J^{AB}$. Their extraction is therefore intrinsically more
sensitive to small variations of these quantities, as discussed below.


The determination of the exchange and anisotropy parameters is carried
out self-consistently. At each temperature, an initial estimate of
$\mathcal K_A$ and $\mathcal K_B$ is used to determine
$\mathcal J^{AB}$, $\mathcal J^{AA}$, and
$\mathcal J^{BB}$ from the finite-$q$ spectrum. The resulting
$\mathcal J^{AB}$ is then combined with the two zone-center
frequencies to update $\mathcal K_A$ and $\mathcal K_B$ as described
above. These two steps are repeated until the interaction parameters
are converged. For the lowest temperature, the initial anisotropies are
taken from the reference scaling
$\mathcal K_r(T)=\mathcal K_r(0)m_r^3(T)$, whereas at subsequent temperatures
the converged solution at the preceding temperature is used as the
initial condition.

The exchange parameters are strongly constrained by the dispersive
part of the spectrum and were found to be insensitive to moderate
variations of the fitting procedure. The anisotropies are more
sensitive because their determination relies on only the two
zone-center frequencies and on the value of the intersublattice
exchange $\mathcal J^{AB}$. In particular, small differences in
$\mathcal J^{AB}$ can lead to appreciable variations in the extracted
$\mathcal K_A$ and $\mathcal K_B$.

In particular, the value of $\mathcal J^{AB}$ obtained from the ASD
fit is optimized to reproduce the finite-$q$ dispersion together with
$\mathcal J^{AA}$ and $\mathcal J^{BB}$. Even though the region
closest to $\Gamma$ is excluded from this fit, small residual
differences between the ASD spectrum and its effective spin-model
representation can be partially accommodated by
$\mathcal J^{AB}$ without appreciably affecting the overall quality
of the fit. Such small variations are largely irrelevant for the
exchange dispersion itself, but become amplified when
$\mathcal J^{AB}$ is subsequently used to separate the much smaller
anisotropy contribution from the zone-center frequencies.

For this reason, we also determine the anisotropies using the
independently calculated Green's-function value
$\mathcal J_{\mathrm{GF}}^{AB}(T)$. Since this quantity is not
obtained by optimizing the ASD finite-$q$ spectrum, it is not affected
by the same parameter compensation and provides a more stable reference
for the decomposition of the zone-center gaps into exchange and
anisotropy contributions. The resulting values are therefore used as
the central ASD-derived anisotropies shown in the main text.

The sensitivity to the choice of intersublattice exchange is quantified
by repeating the extraction with the ASD-fitted value
$\mathcal J_{\mathrm{ASD}}^{AB}(T)$. The difference between the two
determinations is used as a conservative estimate of the systematic
uncertainty,
\begin{equation}
    \Delta\mathcal K_r(T)
    =
    \left|
    \mathcal K_r
    \left[
        \omega_0^\pm(T),
        \mathcal J_{\mathrm{GF}}^{AB}(T)
    \right]
    -
    \mathcal K_r
    \left[
        \omega_0^\pm(T),
        \mathcal J_{\mathrm{ASD}}^{AB}(T)
    \right]
    \right|.
\end{equation}
\textcolor{black}{The resulting $\Delta\mathcal K_r(T)$ values are used to define the
shaded uncertainty bands shown for the ASD-extracted anisotropies in
Fig.~4(a), which therefore quantify the sensitivity
of the extracted $\mathcal K_r(T)$ to the choice of intersublattice
exchange.}

\section{Extraction of the exchange-scaling exponents}
\label{appD}

The low-temperature renormalization of the effective exchange
interactions can be characterized by their scaling with the reduced
sublattice magnetizations. In the absence of transverse spin
correlations, the effective exchange parameters follow the reference
scalings
\begin{equation}
\begin{aligned}
    \mathcal J^{AA}(T)&\propto m_A^2,
    \qquad
    \mathcal J^{BB}(T)\propto m_B^2,
    \\
    \mathcal J^{AB}(T)&\propto m_A m_B=
    \left(\sqrt{m_A m_B}\right)^2 .
\end{aligned}
    \label{eq:SM_uncorrelated_exchange_scaling}
\end{equation}
Thermal spin correlations modify these dependences and can be
expressed as corrections to the corresponding power-law exponents.
The quantities $\varepsilon_{rs}$ shown in the main text quantify
these correlation-induced deviations from the uncorrelated exponent.

\subsection{Extraction from the Green's-function calculations}
\label{subsec:SM_epsilon_extraction}

For each exchange channel, we introduce the
corresponding reduced magnetization variable,
\begin{equation}
m_{AA}=m_A,
\qquad
m_{BB}=m_B,
\qquad
m_{AB}=\sqrt{m_A m_B}.
\label{eq:SM_channel_magnetizations}
\end{equation}
Following the low-temperature scaling analysis introduced in Ref.~\cite{Rzsa2023}, Eqs.~\eqref{eq:eq_JAAm}--\eqref{eq:eq_JBBm} can be written in the common form
\begin{equation}
\frac{\mathcal J^{rs}(T)}{\mathcal J^{rs}(0)}
\sim
m_{rs}^{\beta_{rs}},
\qquad
\varepsilon_{rs}=2-\beta_{rs},
\label{eq:SM_scaling_exponents}
\end{equation}
where $rs=AA,BB,AB$.
The uncorrelated reference corresponds to
$\beta_{rs}=2$, while transverse spin correlations
produce deviations quantified by
$\varepsilon_{rs}$.

The scaling exponents are extracted from the GF
calculations rather than from an additional fit
to the ASD-derived exchange parameters.
The GF and ASD results are in close agreement
in the low-temperature regime considered here,
while the GF calculations avoid the additional
uncertainty associated with extracting the
exchange interactions from the simulated
spin-wave spectra.

For each value of the exchange-partition parameter
$\lambda$, the self-consistent GF equations were
solved on a $100\times100\times100$ reciprocal-space
mesh for 21 temperatures uniformly distributed
between $0$ and $5$~K.

The exponents $\beta_{rs}$ were obtained from
the slopes of
$\ln[\mathcal J^{rs}(T)/\mathcal J^{rs}(0)]$
as a function of $\ln m_{rs}$.
The corresponding correlation corrections
$\varepsilon_{rs}$ then follow directly from
Eq.~\eqref{eq:SM_scaling_exponents}.

\subsection{Exchange-partition model}
\label{subsec:SM_lambda_model}

To investigate the dependence of the correlation
corrections on sublattice inequivalence, we
generalize the exchange-partition model introduced
for two-sublattice antiferromagnets in Ref. ~\cite{Rzsa2023}.

For the rock-salt lattice, the exchange parameters
are defined as
\begin{equation}
\begin{aligned}
12J^{AA}
&=
\frac{2r}{1+r}\lambda J_0,
\quad
12J^{BB}
=
\frac{2}{1+r}\lambda J_0,
\\
6J^{AB}
&=
-(1-\lambda)J_0,
\end{aligned}
\label{eq:SM_exchange_partition}
\end{equation}
Here, $\lambda\in[0,1]$ controls the relative
strengths of the intra- and intersublattice
exchange interactions, while $r= J^{AA} / J^{BB} =3$ sets the
inequivalence between the two sublattices.

The parametrization preserves the average
zero-wave-vector intrasublattice exchange
strength,
\begin{equation}
\frac{J^{AA}_0+J^{BB}_0}{2}
=
\lambda J_0.
\label{eq:SM_average_exchange}
\end{equation}
The limiting cases $\lambda=0$ and $\lambda=1$
correspond to the intersublattice-only and
decoupled-ferromagnet limits, respectively,
as discussed in Sec.~\ref{Low-temperature scaling}.

For this auxiliary ferrimagnetic family, we use
$J_0=60$~meV,
$\mu_A=2\mu_B$,
$\mu_B=3\mu_B$,
and $\alpha_0=1/2$.
A small equal single-ion anisotropy,
$K_A=K_B=0.001$~meV, is included in the numerical GF calculations to regularize the long-wavelength Goldstone mode while
remaining close to the isotropic limit.

\subsection{Symmetric antiferromagnetic reference}
\label{subsec:SM_AFM_reference}

For comparison, we consider the symmetric two-sublattice
antiferromagnetic model of Ref.~\cite{Rzsa2023}, defined by
Eq.~\eqref{eq:27} of the main text. Sublattice equivalence imposes a
single intrasublattice correlation correction,
$\varepsilon_{\mathrm{intra}}^{\mathrm{AFM}}$, and an intersublattice
correction, $\varepsilon_{\mathrm{inter}}^{\mathrm{AFM}}$.
In addition to extracting these quantities numerically from the
self-consistent GF calculations, we evaluate the analytical
low-temperature expressions for the isotropic symmetric system
following Ref.~\cite{Rzsa2023}. These analytical expressions can be written in terms of the normalized geometrical factors of the rock-salt lattice,
\begin{equation}
\begin{aligned}
    \gamma_{\bm q}^{\mathrm{inter}}
    &=
    \frac{\cos q_xa+\cos q_ya+\cos q_za}{3},
    \\
    \gamma_{\bm q}^{\mathrm{intra}}
    &=
    \frac{
        \cos q_xa\cos q_ya+
        \cos q_xa\cos q_za+
        \cos q_ya\cos q_za
    }{3},
    \label{eq:SM_gamma_factors}
\end{aligned}
\end{equation}
with
$\gamma_{\bm 0}^{\mathrm{inter}}
=\gamma_{\bm 0}^{\mathrm{intra}}=1$.
We further define
\begin{equation}
\begin{aligned}
    D_{\bm q}(\lambda)
    &=
    \left[
        1-\lambda\gamma_{\bm q}^{\mathrm{intra}}
    \right]^2
    -
    (1-\lambda)^2
    \left(
        \gamma_{\bm q}^{\mathrm{inter}}
    \right)^2,
    \\
    \mathcal N(\lambda)
    &=
    \left\langle
    \frac{
        1-\lambda\gamma_{\bm q}^{\mathrm{intra}}
    }{
        D_{\bm q}(\lambda)
    }
    \right\rangle_{\bm q},
    \label{eq:SM_Dq_normalization}
\end{aligned}
\end{equation}
Here, $\langle\cdots\rangle_{\bm q}$ denotes the Brillouin-zone average. The correlation corrections are then given by
\begin{equation}
\begin{aligned}
    \varepsilon_{\mathrm{inter}}^{\mathrm{AFM}}(\lambda)
    &=
    \frac{1}{\mathcal N(\lambda)}
    \left\langle
    \frac{
        (1-\lambda)
        \left(\gamma_{\bm q}^{\mathrm{inter}}\right)^2
    }{
        D_{\bm q}(\lambda)
    }
    \right\rangle_{\bm q},
    \label{eq:SM_epsilon_AFM_1}
\end{aligned}
\end{equation}

\begin{equation}
\begin{aligned}
    \varepsilon_{\mathrm{intra}}^{\mathrm{AFM}}(\lambda)
    &=
    \frac{1}{\mathcal N(\lambda)}
    \left\langle
    \frac{
        \gamma_{\bm q}^{\mathrm{intra}}
        \left[
            1-\lambda\gamma_{\bm q}^{\mathrm{intra}}
        \right]
    }{
        D_{\bm q}(\lambda)
    }
    \right\rangle_{\bm q}.
    \label{eq:SM_epsilon_AFM_2}
\end{aligned}
\end{equation}

At $\bm q=\bm 0$, the denominator $D_{\bm q}(\lambda)$ vanishes due to the Goldstone mode of the isotropic reference system. For $0\leq\lambda<1$, $D_{\bm q}(\lambda)\propto q^2$
as $q\to0$. At $\lambda=1$, the denominator vanishes as $q^4$, but the numerator
of the normalization integrand simultaneously vanishes as $q^2$. Consequently, the integrands entering $\mathcal N(\lambda)$ and
Eq.~\eqref{eq:SM_epsilon_AFM_1}--\eqref{eq:SM_epsilon_AFM_2} exhibit at most an integrable $1/q^2$ singularity in three dimensions, and the Brillouin-zone averages remain finite in the thermodynamic limit. In the numerical evaluation on a discrete reciprocal-space mesh, the isolated $\bm q=\bm 0$ point is therefore omitted. Because of this long-wavelength behavior, the
discrete Brillouin-zone sums converge slowly with the reciprocal-space
mesh size. To remove this finite-mesh effect, the analytical expressions in
Eq.~\eqref{eq:SM_epsilon_AFM_1}--\eqref{eq:SM_epsilon_AFM_2} were evaluated for several
$N\times N\times N$ meshes and extrapolated to the thermodynamic limit
using the leading finite-size form
\begin{equation}
    \varepsilon_{rs}(N,\lambda)
    =
    \varepsilon_{rs}(\infty,\lambda)
    +
    \frac{c_{rs}(\lambda)}{N}.
    \label{eq:SM_epsilon_finite_size}
\end{equation}
The extrapolation was performed using
$N=64,80,100,128,160,200,$ and $256$.

This procedure yields
$\varepsilon_{\mathrm{inter}}^{\mathrm{AFM}}\to0.341$
for $\lambda\to0$ and
$\varepsilon_{\mathrm{intra}}^{\mathrm{AFM}}\to0.255$
for $\lambda\to1$, consistently with the simple-cubic and fcc
ferromagnetic limits, respectively. The solid gray curves in
Fig.~\ref{fig:scaling} correspond to these thermodynamic-limit
analytical results, whereas the gray symbols are obtained independently
from the finite-anisotropy self-consistent GF calculations. The overall agreement between the two approaches, together with the
recovery of the simple-cubic and fcc limiting values, provides a
consistency check of the numerical extraction of the scaling exponents.

\bibliography{Ref}

\end{document}